\pdfoutput=1
\documentclass{aa2}

% ...

\usepackage{amsmath, amssymb}
\usepackage[english=british]{csquotes}
\usepackage[T1]{fontenc}
\usepackage[utf8]{inputenc}
\usepackage{lmodern}
\usepackage{graphicx}
\usepackage[dvipsnames]{xcolor}
\usepackage{natbib}
 \bibpunct{(}{)}{;}{a}{}{,}
\usepackage{hyperref}
\usepackage{verbatim}
\usepackage[capitalize]{cleveref}

\usepackage{physics} %For equation type setting
\usepackage[alsoload=astro]{siunitx} %For SI Units

\newcommand{\Omm}{\Omega_\mathrm{m}}

\newcommand{\dgm}{\ensuremath{\operatorname{Dgm}}}
\newcommand{\mbR}{\ensuremath{\mathbb{R}}}

\newcommand{\Map}{\mathcal{M}_\mathrm{ap}}

\makeatletter
\renewcommand*\aa@pageof{page \thepage{} of \pageref*{LastPage}}
\makeatother

\begin{document}

\title{
Persistent homology in cosmic shear II: A tomographic analysis of DES-Y1
}
\titlerunning{Persistent homology in cosmic shear II}

\author{Sven Heydenreich\inst{1}, Benjamin Br\"uck\inst{2}, Pierre Burger\inst{1}, Joachim Harnois-D\'{e}raps\inst{3,4}, Sandra Unruh\inst{1,5},
Tiago Castro\inst{6,7,8}, Klaus Dolag\inst{9,10}, Nicolas Martinet\inst{11}
      }
\authorrunning{Heydenreich et al.}

\institute{
      Argelander-Institut f\"ur Astronomie, Auf dem H\"ugel 71, 53121 Bonn, Germany 
      \and 
      ETH Z\"urich, Department of Mathematical Sciences, R\"amistrasse 101, 8092 Z\"urich, Switzerland
      \and
      School of Mathematics, Statistics and Physics, Newcastle University, Herschel Building, NE1 7RU, Newcastle-upon-Tyne, UK
      \and
      Astrophysics Research Institute, Liverpool John Moores University, 146 Brownlow Hill, Liverpool L3 5RF
      \and
      Ruhr University Bochum, Faculty of Physics and Astronomy, Astronomical Institute (AIRUB), German Centre for Cosmological Lensing, 44780 Bochum, Germany
      \and
      Osservatorio Astronomico di Trieste, via Tiepolo 11, I-34131, Trieste, Italy
      \and
      Institute for Fundamental Physics of the Universe, via Beirut 2, 34151, Trieste, Italy
      \and 
      INFN – Sezione di Trieste, I-34100 Trieste, Italy  
      \and
      Universitäts-Sternwarte, Fakultät für Physik, Ludwig-Maximilians-Universität München, Scheinerstr.1, 81679 München, Germany 
      \and
      Max-Planck-Institut für Astrophysik, Karl-Schwarzschild-Straße 1, 85741 Garching, Germany
      \and
      Aix-Marseille Univ, CNRS, CNES, LAM, Marseille, France
      \\ \email{sven@astro.uni-bonn.de}
      }

\date{Version \today; received xxx, accepted yyy} % day month year

\abstract
{
We demonstrate how to use persistent homology for cosmological parameter inference in a tomographic cosmic shear survey. We obtain the first cosmological parameter constraints from persistent homology by applying our method to the first-year data of the Dark Energy Survey.

To obtain these constraints, we analyse the topological structure of the matter distribution by extracting persistence diagrams from signal-to-noise maps of aperture masses. This presents a natural extension to the widely used peak count statistics. Extracting the persistence diagrams from the {\it cosmo}-SLICS, a suite of $N$-body simulations with variable cosmological parameters, we interpolate the signal using Gaussian Processes and marginalise over the most relevant systematic effects, including intrinsic alignments and baryonic effects.

We find for the structure growth parameter $S_8=0.747^{+0.025}_{-0.031}$, which is in full agreement with other late-time probes. %and confirms the tension with analyses of the Cosmic Microwave Background. 
We also constrain the intrinsic alignment parameter to $A=1.54\pm 0.52$, ruling out the case of no intrinsic alignments at a $3\sigma$-level.

%statistical properties of the Universe -> cosmology. topological structure analysis of observable weak lensing shear -> cosmic shear/aperture mass and redshift. persistence diagrams / persistent homology extends info from commonly known/used peak statistics. analysis pipeline/approach from \citet{heydenreich2021} plus different summary statistics plus cross-bin tomographic analysis plus a bunch of systematics plus synergy with 2pt correlations. some results for $S_8 $, maybe $w_0$, w/ \& w/o 2pcf. some outrageous claim about ability breaking degeneracies / solving the $S_8$ tension.
}


\maketitle

\section{Introduction}
\label{sec:introduction}
In the past decades, weak gravitational lensing has emerged as an indispensable tool to study the large-scale structure (LSS) of the Universe. Weak lensing primarily relies on accurate shape and distance measurements of galaxies. Ongoing and recently completed surveys have provided the community with a sizeable amount of high-quality data; e.g., the Kilo Degree Survey \citep[KiDS,][]{dejong2013}, the Dark Energy Survey \citep[DES,][]{flaugher2005}, and the Hyper Suprime-Cam Subaru Strategic Program \citep[HSC,][]{aihara2018}. Further surveys are scheduled to start observing in the next years; they will probe deeper and larger areas enabling measurements of cosmological parameters with sub-percent accuracy; e.g., the Vera Rubin Observatory's Legacy Survey of Space and Time \citep[LSST,][]{ivezic2008}, the {\it Euclid} survey \citep{laureijs2011}, and the Nancy Grace Roman Space Telescope survey \citep[RST,][]{spergel2013}. These upcoming surveys are of special relevance for solving tensions related to measurements of the structure growth parameter $S_8= \sigma_8 \sqrt{\Omega_\mathrm{m}/0.3}$ \citep{hildebrandt2017, planck2020, joudaki2020, heymans2021,DES2021} which is defined along the main degeneracy direction in conventional weak lensing studies. Here, $\Omega_\mathrm{m}$ is the dimensionless matter density parameter and $\sigma_8$ parametrises the amplitude of the matter power spectrum. Improved data and independent analysis choices are crucial to determine whether this tension is due to new physics, a statistical fluctuation, or the manifestation of unknown systematics. For example, \citet{Joudaki:2017} showed that the current tension in $S_8$ between the CMB and the local Universe can be lifted when allowing for a dynamical dark energy model, meaning that measuring the equation-of-state of dark energy is of utmost importance in the next decades.

Shear 2-point statistics have emerged as the prime analysis choice for cosmic shear as they present a number of key advantages \citep[e.g.][]{DESY3_Secco, hikage2019, asgari2021}. Such statistics are physically motivated by the fact that they describe the early Universe almost perfectly. The late Universe, however, contains a considerable amount of non-Gaussian information that is not captured by 2-point statistics, such that jointly investigating second- and higher-order statistics increases the constraining power on cosmological parameters \citep[see, e.g.][]{Berge2010,Pyne2021}. This additional information is currently explored with a variety of analysis tools, which either use analytical models \citep[e.g.,][]{Halder2021,Burger2021} or rely on large suites of numerical simulations. Relevant examples for simulation-based analysis are peak statistics \citep[][and references therein, hereafter M+21]{martinet2021}, and Minkowski functionals \citep[e.g.,][]{shirasaki2014, petri2015, parroni2020} which are both based on aperture mass maps constructed from shear fields. M+21 also showed that a joint analysis of peaks and 2-point correlation functions (2PCF) improves cosmological constraints on $S_8$, $\Omega_\mathrm{m}$, and the Dark Energy equation-of-state parameter $w_0$ by 46\%, 57\% and 68\%, respectively. \citet{zurcher2021} showed that a joint analysis using 2PCF with Minkowski functionals, a topological summary statistic, on aperture mass maps increases the figure-of-merit in the {$\Omega_\mathrm{m}$--$\sigma_8$} plane by a factor of 2.

In this paper, we focus on persistent homology, a topological method that combines the advantages of peak statistics and Minkowski functionals but also captures information about the environment of topological features. Persistent homology specialises in recognising persistent topological structures in data and we refer the interested reader to a recent review written by \citet{wassermann2018} who highlights its diverse applications in various fields. Following early concepts about persistent homology and Betti numbers in cosmology \citep{weygeart2011}, several groups have formalised the approach \citep{sousbie2011, pranav2017, feldbrugge2019, pranav2021}. In particular, \citet{kimura2017} were the first to show that the hierarchical topological structure of the galaxy distribution decreases with increasing redshift using small patches of Sloan Digital Sky Survey (SDSS). 
More recently, \citet{xu2019} developed an effective cosmic void finder based on persistent homology, while \citet{kono2020} detected baryonic acoustic oscillations in the quasar sample from the extended Baryon Oscillation Spectroscopic Survey in SDSS. Moreover, \citet{biagetti2020,Biagetti:2022} showed with simulations that persistent homology is able to identify primordial non-Gaussian features. \citet[][hereafter H+21]{heydenreich2021} performed a mock analysis using persistent homology on cosmic shear simulations, highlighting its potential to break the degeneracy between $S_8$ and $w_0$. 

Persistent homology summarises the topological structure of data in so-called persistence diagrams. There are different methods for performing statistical analyses on such diagrams (see Sect.~\ref{sect:pers_stats}). In H+21, we worked with persistent Betti numbers. In this work, we opted for `heatmaps', which constitute a more robust statistic for persistence diagrams.
We extract heatmaps from a series of mock data that match the DES-Y1 survey properties \citep[][hereafter HD+21]{flaugher2005,harnois-deraps:2021}, including a {\it Cosmology Training Set}, a {\it Covariance Training Set} and a suite of {\it Systematics Training Set}, constructed from the SLICS \citep{harnois-deraps2015}, the {\it cosmo}-SLICS simulations \citep{harnois-deraps2019} and the {\it Magneticum} hydrodynamical simulations \citep{Biffi:2013,Saro:2014,Steinborn:2015,Steinborn:2016,Dolag:2015,Teklu:2015,Bocquet:2016,Remus:2017,Castro:2018,MagneticumBox2b}. Following HD+21, we then train a Gaussian process regression (GPR) emulator, which is fed to a Markov Chain Monte Carlo (MCMC) sampler to obtain cosmological parameter estimates. We significantly expand on the results from H+21 by including the main systematic effects related to cosmic shear analyses, namely photometric redshift uncertainty, shear calibration, intrinsic alignment of galaxies, baryon feedback and masking. These systematics, in particular baryon feedback and intrinsic alignments, account for $25\%$ of our reported final error budget. Furthermore, as introduced in M+21, our results are obtained for a tomographic topological data analysis where we include the cross-redshift bins analyses. This leads to the first cosmological parameter constraints obtained from persistent homology based on analysing cosmic shear data, here provided by the DES year-1 survey \citep{DESY1_data}.

The paper is organised as follows: In Sect.~\ref{sec:data} we describe the data and simulations; the theoretical background on persistent homology, a description of our data compression methods, the formalism for the 2-point statistics and the cosmological parameter estimation are presented in Sect.~\ref{sect:background}. In Sect~\ref{sec:systematics} we discuss our mitigation strategies for systematic effects and show the validation of our pipeline in Sect.~\ref{sect:validation}. We finalise our work with the results shown in Sect.~\ref{sect:results} and our discussion in Sect.~\ref{sect:discussion}.

\section{Data and Simulations}
\label{sec:data}
\subsection{DES-Y1 data}
\label{subsec:data}
We use in this work the public\footnote{DES-Y1 catalogues: des.ncsa.illinois.edu/releases/dr1} Year-1 data  released by the Dark Energy Survey presented in \citet[][DES-Y1 hereafter]{DESY1_data}. The primary weak lensing data consist of a galaxy catalogue in which positions and ellipticities are recorded for tens of millions of objects, based on observations from DECam mounted at the Blanco telescope at the Cerro Tololo Inter-American Observatory \citep{DES_CAM}. The galaxies selected in this work match those of \citet[][hereafter T+18]{troxel2018} and HD+21, applying the {\sc flags select}, {\sc Metacal}, and the {\sc redmagic} filters to the public catalogues, yielding a total unmasked area of 1321 deg$^2$ and 26 million galaxies.

The shear signal $\gamma_{1/2}$ is inferred from the {\sc Metacalibration} technique \citep{METACAL}, which further provides each galaxy with a {\sc Metacal} response function $S_i$ that must be included in the measurement. As explained in T+18, this method requires a prior on an overall multiplicative shear correction of $m \pm \sigma_m = 0.012 \pm 0.023$, which we then use to calibrate the measured galaxy ellipticities as $\epsilon_{1/2} \rightarrow \epsilon_{1/2}(1+m)$. We then assume that these ellipticities are an unbiased estimator for the shear $\gamma$.

Following T+18, the galaxy sample is further split into four tomographic bins based on their individual estimated photometric redshift $Z_B$, which is measured with the {\sc bpz} method \citep{BPZ}. At this point, the redshift distribution of the four tomographic populations are estimated with the `DIR'\footnote{This method relies on the direct calibration of the $n(z)$ from a sub-sample of DES-Y1 galaxies for which external spectroscopic data are available. See \citet{Lima} for more details.} method, following \citet{joudaki2020} and HD+21. As argued in these two references,
the DIR approach is more robust to potential residual selection effects  in their training  sample,
compared to the DES-Y1 {\sc bpz} stacking method presented in \citet{DESY1_redshifts}.  Although it could also be affected by
incomplete spectroscopy and colour pre-selection \citep{GruenBrimioulle},
bootstrap resampling of the spectroscopic samples points towards a significantly smaller uncertainty in the mean redshift of the populations, achieving $\sigma_z=0.008, 0.014, 0.011\, \text{and}\, 0.009$ for tomographic bins 1..4, respectively \citep{joudaki2020}. Despite these important differences, the DIR $n(z)$ is consistent with the fiducial  estimate presented by the DES collaboration. It brings in excellent agreement the DES-Y1 and the KV-450 cosmic shear data \citep[][also based on the DIR method]{KV450}, and it was further shown in \citet{Asgari:2020} that the inferred $S_8$ value is affected by less than $1\sigma$.

\subsection{Mock galaxy catalogues}
\label{subsec:mocks}

The analysis presented in this work largely follows the simulation-based inference methods of HD+21, which completely relies on numerical weak lensing simulations for the cosmology inference, the estimation of the uncertainty and the mitigation of systematics and secondary effects. Most of the mock data used in this work has been presented in HD+21, which we review in this Section.  

\begin{itemize}
\item {\it Cosmology Training Set:} This set of simulations is used to model the dependence of the signal on cosmology. Based on the {\it cosmo}-SLICS \citep{harnois-deraps2019}, it consists of weak lensing light-cones sampling 26 points in a $w$CDM cosmological model (i.e. cold dark matter with dark energy beyond the cosmological constant $\Lambda$), where 25 points are distributed on a Latin Hypercube, covering the ranges $\Omega_{\rm m} \in [0.10,0.55]$, $S_8 \in [0.60, 0.90]$, $h \in [0.6, 0.82]$ and $w_0 \in [-2.0, -0.5]$, where $h$ is the reduced Hubble parameter. The last point is set manually to a fiducial, $\Lambda$CDM cosmology. Each node consists of two independent $N$-body simulations produced by {\sc cubep$^3$m} \citep{CUBEP3M}, with initial conditions designed such as to suppress the sampling variance. The code follows the non-linear evolution of 1536$^3$ particles in a 505 $h^{-1}$ Mpc box, producing between 15 and 28 mass sheets of co-moving thickness equivalent to half the box size, filling up a $10\times 10\,\mathrm{deg}^2$ light-cone to $z=3.0$. Random orientations and shifting are introduced in this process such that a total of 25  {\it pseudo}-independent light-cones are generated, per $N$-body run. Five of these are used in the current paper, out of 25, which is sufficient to model the statistics within the DES-Y1 precision (as in HD+21). Validation tests revealed that the third line-of-sight from the first $N$-body seed is a statistical outlier: for example the standard deviation of the convergence $\sigma_\kappa$ differs from the mean of the full {\it cosmo}-SLICS light-cones by more than $4\sigma$. Due to the limited size of our training sample, this particular line-of-sight could bias our cosmological model. We thus skipped over it, and verified afterwards that the results are not strongly affected by this choice, although it slightly improves the accuracy on $S_8$ during the validation test. In total, the  {\it Cosmology Training Set} consists of $26\times9 =234$ survey realisations.

\item {\it Covariance Training Set:} This suite is mainly used to estimate the sampling covariance in the data vector. Based on the SLICS \citep{harnois-deraps2015}, it is produced from 124 fully independent $N$-body realisations, with the same  mass resolution and simulation volume as the {\it cosmo}-SLICS. All carried out at the same cosmology, these light-cones started from different noise realisations of the initial conditions, thereby sampling the statistical variance in the data. The mean over all measurements from the {\it Covariance Training Set} is also independent from the  {\it Cosmology Training Set} and well converged towards the ensemble average, making this an ideal data set with which we validate our cosmology inference pipeline later on. 
\item {\it Systematics Training Set - Mass resolution:} The force resolution of $N$-body simulations is limited by the number of particles, the choice of softening length and the force accuracy setting. This inevitably translates into  a decrease in the clustering of dark matter in the highly non-linear scales, which in turn affects the statistics under study. The SLICS-HR are a suite of high-resolution simulations introduced in \citet{harnois-deraps2015}, in which the force accuracy of {\sc cubep$^3$m} has been significantly increased, yielding 5 light-cones with more accurate mass densities. As detailed in Sect.~\ref{sec:systematics}, we verify that our training data are not strongly affected by this known limitation. %force resolution, and nevertheless employ a correction factor to calibrate our model for the analysis of real data.
%vector the elements for which the discrepancies between these and the {\it Cosmology Training Set} are large. %\textcolor{purple}{[Do we do that, or just made sure that the cosmology inference run on these is unbiased? SH: We boost our data vector prediction (for the DES-Y1 data) by the ratio HR/no-HR. We also check that this ratio is $\ll 1\sigma$, which it is for all data points.]}
\item {\it Systematics Training Set - Baryons:} Baryonic feedback processes from sustained stellar winds, supernovae and active galactic nuclei are known to redistribute the matter around over-dense regions of the Universe, in a manner that directly affects the weak lensing measurements \citep{Semboloni11}. If left unmodelled, these processes will significantly bias the inferred cosmology in analyses based on 2PCF or non-Gaussian statistics \citep[e.g.][]{MinimaPeaks, zurcher2021,martinet21b}. In this work, our approach consists of measuring our statistics in hydrodynamical simulations in which the baryon feedback can be turned on and off.  The relative impact on the data vector is then used to model the effect of baryons on our statistics. As in HD+21, we use the Magneticum simulations\footnote{www.magneticum.org} to achieve this, more precisely the  Magneticum {\it Run}-2 and {\it Run}-2b \citep{MagneticumBox2b}, in which stellar formation, radiative cooling, supernovae and AGN feedback are implemented in  cosmological volumes of 352 and 640 $h^{-1}$ Mpc, respectively, with a spatial resolution that is high enough to capture the baryonic effects at scales relevant to our study. The adopted cosmology is consistent with the SLICS cosmology, with $\Omega_{\rm m}$ = 0.272, $h$ = 0.704, $\Omega_{\rm b}$ = 0.0451, $n_{\rm s}$ = 0.963, and $\sigma_8$ = 0.809. These simulations reproduce a number of key observations, including many statistical properties of the large-scale, intergalactic, and intercluster medium \citep[see][for more details]{2014MNRAS.442.2304H, Teklu:2015, LensingPDF_baryons}. Moreover, the resulting overall feedback is consistent with that of the BAHAMAS simulations \citep{BAHAMAS}, which are based on a completely independent sub-grid calibration method. The  {\it Baryons Training Set} and their dark-matter-only counterpart  are used to inspect the impact of baryonic physics on the data vector,  from which we extract a correction factor used to forward-model the effect on dark-matter only simulations.  Full details on the treatment of the systematics are presented in Section  \ref{sec:systematics}.
\item {\it Systematics Training Set - Photometric redshifts:} The redshift distribution of the data is known to a high precision within the DIR method, however the residual uncertainty must be accounted for in the analysis. For this, we use the mocks described in HD+21 in which the $n(z)$ has been shifted by a small amount, in order to study the impact on the signal. These sample at ten points the posterior of the expected shifts in the mean redshifts of the DIR method itself \citep{joudaki2020}, and in each case we construct 10 full survey realisations at the {\it cosmo}-SLICS fiducial cosmology, from which we extract our statistics. This approach allows us to measure the derivative of the persistent homology statistics with respect to shifts in d$z$. The priors on d$z$ are listed in Table \ref{tab:priors}.
%\item {\it Systematics Training Set - Shape calibration:}

\item {\it Systematics Training Set - Intrinsic Alignments:}
The assumption that the observed shapes of galaxies are randomly aligned in absence of foreground lensing matter fails to account for their intrinsic alignment (IA), an important contribution that arises from a coupling between their shapes and the large scale structure they are part of \citep[for a review see][]{Joachimi_IA_review}. This important secondary signal tends to counteract  the cosmic shear signal, which can therefore interfere in the cosmological inference. Although there exist analytical models to describe this effect for two-point functions, higher-order statistics must rely on IA-infused  simulations to account for this important effect. In this work we use the infusion method presented in \citet{JHD21b}, where intrinsic galaxy shapes are linearly coupled with the projected tidal field, consistent with the non-linear alignment model of \citet{BridleKing}. Although the redshift distribution of these mocks exactly follow that of the data, their construction requires that the galaxy ellipticities linearly trace the simulation density fields, whose positions therefore no longer replicate that of the DES-Y1 data. These mocks have no masking nor {\sc Metacalibration} responses included, and are therefore used to estimate the relative impact of IA on our persistent homology measurements. They have an IA amplitude that is allowed to vary,  as controlled by the $A_{\rm IA}$ parameter. We measure the persistent homology statistics from 50 {\it cosmo}-SLICS light-cones at the fiducial cosmology, for values of $A_{\rm IA} \in [-5.0 , 5.0]$, and use these to construct a derivative, similar to the way we handle the photometric redshift uncertainty. 
\end{itemize}

The output of each simulation is a series of 100 deg$^2$ lensing planes  that serve to assign convergence ($\kappa$) and shear ($\gamma_{1/2}$) to copies of the DES-Y1 data. As described in HD+21, the survey footprint is segmented into 19 regions, or tiles, which all fit inside our simulated maps. The summary statistics are computed individually on each tile and combined afterwards to construct the data vector. In this construct, the galaxy positions,  ellipticities $\epsilon_{1/2}$ and {\sc Metacalibration} weights $S_i$ in the mock data exactly match  that of the real data, avoiding possible biases arising in non-Gaussian statistics when these differ \citep[see e.g.][Appendix D of H+21]{Kacprzak:2016}.
Mock ellipticities are obtained by rotating the observed ellipticities by a random angle, and combining the resulting randomised signal ${\boldsymbol \epsilon}_{\rm n}$ with the simulated (noise-free) reduced shear $\boldsymbol{g}$ via:  
\begin{equation}
{\boldsymbol \epsilon} = \frac{{\boldsymbol \epsilon}_{\rm n} + {\boldsymbol g}}{ 1 + {\boldsymbol \epsilon}_{\rm n}{\boldsymbol g}^*}\; ,
 \label{eq:ellipticity_definition}
\end{equation}
where bold symbols denote complex numbers (e.g. ${\boldsymbol g} = g_1 + {\rm i}g_2$). We calculate the reduced shear as ${\boldsymbol g}={\boldsymbol \gamma}/(1-\kappa)$.
In total we compute 10 shape-noise realisations for every simulated survey realisation, each using a different random seed in the rotation. This procedure allows us to average out a large part of the fluctuations that are introduced by the shape noise, improving both our predictions and our estimate of the sample covariance, while preserving exactly the data noise levels.

Redshifts are assigned to every simulated galaxy by sampling from the DIR redshift distribution corresponding to the tomographic bin they belong to.

\subsection{Calculating maps of aperture masses}
\label{sec_calculating_aperture_mass}
As in H+21, we perform our computations on signal-to-noise maps of aperture masses \citep{Schneider:1996,bartelmann:2001}, computing the signal $\Map$(\vec{\theta})  and noise $\sigma\left(\Map(\vec{\theta})\right)$ on a grid as:
\begin{align}
    \Map(\vec{\theta}) = & \frac{1}{n_\mathrm{gal}\sum_i w_iS_i}\sum_i Q(|\vec{\theta}_i-\vec{\theta}|)w_i\epsilon_\mathrm{t}(\vec{\theta}_i;\vec{\theta})\; , \label{eq:map_def}\\
    \sigma\left(\Map(\vec{\theta})\right) = & \frac{1}{\sqrt{2}n_\mathrm{gal}\sum_i w_iS_i} \sqrt{\sum_i|w_i\epsilon(\vec{\theta}_i)|^2Q^2(|\vec{\theta}_i-\vec{\theta}|)}\; ,
    \label{eq:map_noise_def}
\end{align}
where the $w_i$ are optional weights assigned to measured galaxy ellipticities (set to 1.0 in this work), $S_i$ are the respective responses calculated by the \textsc{Metacalibration} shear estimator (T+18), and the tangential component of the shear $\epsilon_\mathrm{t}(\vec{\theta}_i;\vec{\theta})$ is calculated via 
\begin{equation}
    \epsilon_\mathrm{t}(\vec{\theta}_i;\vec{\theta}) = -(\epsilon_1+\mathrm{i}\epsilon_2)\frac{(\vec{\theta}_i-\vec{\theta})^*}{(\vec{\theta}_i-\vec{\theta})} \, .
\end{equation}
We then compute the signal-to-noise map (S/N-map) of aperture masses as the ratio between the two quantities. As before, we use the following Q-filter function \citep[][hereafter M+18]{Schirmer:2007,Martinet:2018}:
\begin{align}
  Q(\theta) = & \left[1 + \exp \left(6 -150 \frac{\theta}{\theta_{\rm ap}}\right) + \exp \left(-47 +50 \frac{\theta}{\theta_{\rm ap}}\right)\right]^{-1}
             \nonumber\\
             & \quad \times \left(\frac{\theta}{x_{\rm c}\theta_{\rm ap}}\right)^{-1} \tanh \left(\frac{\theta}{x_{\rm c}\theta_{\rm ap}}\right)\, .
             \label{eq:filterfunction}
\end{align}
with a concentration index of $x_c=0.15$ \citep{Hetterscheidt:2005}, which was chosen to optimally select the mass profiles of dark matter halos \citep{Navarro:1997}. For the filter radius we choose $\theta_\mathrm{ap} = 12.5\arcmin$. As in H+21, we compute the S/N maps by distributing both galaxy ellipticities $\epsilon_i$ and their squared moduli $|\epsilon_i|^2$ for each tile on a $(600\times 600)$ pixel grid, and perform the convolutions in Eqs.~\eqref{eq:map_def} and \eqref{eq:map_noise_def} via a Fast Fourier-Transform. Contrary to previous work, we use a cloud-in-cell algorithm to distribute the galaxy ellipticities on a grid, yielding more accurate results for small scales when dealing with high-quality data.

%We perform our analysis on tomographic mock data of the DES-Y1 survey. 
As shown in M+21, the traditional approach of computing the aperture mass statistics for individual tomographic bins only (hereafter auto-bins) does not yield optimal results. Instead we perform the computation for all combinations of tomographic bins by concatenating the respective galaxy catalogues (cross-bins). 
This approach allows us to extract additional information about correlated structure along the line-of-sight. For example, a massive, nearby galaxy cluster can be detected as a peak in the S/N maps for tomographic bins 1 and 2. However, if we were to only analyse persistence heatmaps of the two respective bins, both would register the cluster as a peak, but the information that the peak is at the same position in both maps would be lost. To utilise this information, we also need to analyse the S/N map of a combination of both bins.
%While this leads to a total of 15 tomographic redshift bins that need to be analysed, we make sure that we do not miss any information residing in combinations of redshift bins.
Based on the four fiducial DES-Y1 redshift bins, this optimised method leads to 15 bin combinations (1, 2, 3, 4, 1$\cup$2, ..., 3$\cup$4, ..., 1$\cup$2$\cup$3$\cup$4) from which we extract heatmaps.

As the galaxies in our mock data follow the exact positions of the real galaxy catalogue, they are subject to the overall survey footprint and internal masked regions.
%As the mock data follows the footprint of the DES-Y1 survey, it is subject to gaps and masks.
We only want to consider the parts of our S/N maps where we have sufficient information from surrounding galaxies, therefore we construct our own mask in the following way: We combine the galaxy catalogues of all four tomographic bins and distribute these galaxies on a grid. Then we mask all pixels of the tile where the effective area containing galaxies within the aperture radius $\theta_\mathrm{ap}$ is less than 50\%. In particular, we mask the boundary of each tile to ensure that neighbouring tiles are treated as independent in the persistence calculations (compare H+21). This mask is then applied to every combination of tomographic bins of the respective survey tile.

%As can be seen in App.~D of H+21, the distribution of shape noise not only affects the sample covariance, but also the data itself. \textcolor{purple}{Since our simulations correctly capture the shot noise and shape noise in the data, we are protected against possible }

%To accurately trace the shape noise of the DES-Y1 survey, we rotate the \textcolor{purple}{observed galaxy} ellipticities $\epsilon_\mathrm{obs}$ by a random angle to simulate a \textcolor{purple}{pure noise} ellipticity $\epsilon_n$ and add these to the shear extracted from simulations $g$ via
%\begin{equation}
%\epsilon = \frac{\epsilon_{\rm n} + g}{ 1 + \epsilon_{\rm n}g^*}\; ,
% \label{eq:ellipticity_definition}
%\end{equation}
%where we assume for the reduced shear $g=\gamma/(1-\kappa)\approx\gamma$ due to the fact that $\kappa \ll 1$ almost everywhere.
%In total we compute 10 shape-noise realisations \textcolor{purple}{for} every light-cone, each one \textcolor{purple}{using a different random seed in the rotation}. This procedure allows us to average out a large part of the fluctuations that are introduced by the shape noise, improving both our predictions and our estimate of the sample covariance.

With this method we compute the S/N maps for each of the 19 tiles of the DES-Y1 survey footprint, for each of the 15 tomographic bins and for each of the 10 shape noise realisations. The next section describes how cosmological information is extracted from these maps with statistics based on  persistent homology.  %and concatenate these maps, yielding a \textcolor{purple}{single} S/N map for the whole survey. \textcolor{purple}{[JHD: Not sure I'd say this, since the tiles are effectively analyse independently. Don't you think it could confuse the reader?]}

\section{Methods}
\label{sect:background}
We use methods from persistent homology to quantify the statistical properties of S/N maps of aperture masses and analyse their dependence on the underlying cosmological parameters. The main idea can be described as follows:

We take a S/N map of aperture masses and apply a threshold to that map. We then cut off all parts where the value of the S/N map exceeds that threshold (compare Fig.~\ref{fig:aperture_mass_map_with_cutoff}). This gives rise to two types of \emph{topological features}. The first type are connected components, i.e.~ regions of low S/N that are surrounded by a region of higher S/N, which is above the cut-off threshold. These connected components correspond to local minima in the S/N map, which in turn correspond to an underdensity in the matter distribution. The second type of topological features are holes, i.e.~regions of high S/N that are above the cut-off threshold, with an environment of S/N that surrounds them and is lower than the cut-off threshold. These holes correspond to local maxima of the S/N map, which indicate an overdensity in the underlying matter distribution.

When the cut-off threshold is gradually increased, these features change. Connected components start to show up (are \emph{born}) once the threshold is higher than their minimum S/N value. At some higher threshold, the connected component will merge with a different connected component (or \emph{die})\footnote{When two connected components merge, the one that was born at a lower threshold survives. This is known as the \textit{elder rule}.}. Similarly, an overdensity starts to form a hole once the cut-off threshold exceeds the S/N value of its environment. This hole is completely filled in once the threshold exceeds the maximal S/N value of the overdensity.

For each such topological feature, we write $b$ for its birth (the threshold at which it is born) and $d$ for its death (the threshold at which it dies). We plot the collection of all points $(b,d)$ as a scatter plot, called the \emph{persistence diagram} $\dgm$; we write $\dgm_0$ for the persistence diagram of connected components  and $\dgm_1$ for the one of the holes  (see Fig.~\ref{fig:persistence_diagram}). In particular, it is straight-forward to recover the peak count statistics from this: The death of a hole corresponds to the maximal S/N value of an overdensity, so the set of deaths is the collection of all peaks in the S/N map. However, persistent homology offers one crucial advantage: The \emph{persistence} of a feature, defined by $d-b$, yields information about how much a peak protrudes from its surrounding environment. In particular, features with a very small persistence are more likely to be caused by noise fluctuations, which can be taken into account in the following statistical analysis. Persistent homology offers a natural way to account for masked regions, which we describe in the next subsection. We denote the persistence diagrams that account for the presence of masked regions by $\dgm^M_0$ and $\dgm^M_1$.

From the persistence diagrams $\dgm_0^M$ and $\dgm_1^M$ we then create so-called \emph{heatmaps} by smoothing the diagrams with a Gaussian. Every point of the heatmap can now be used for a statistical analysis of the persistent topological structure of the S/N maps of aperture mass.

In the next two subsections, we will give a slightly more formal introduction into these statistics derived from persistent homology and describe their application.

\subsection{Persistent homology}
\begin{figure*}
    \sidecaption
    \includegraphics[width=12cm]{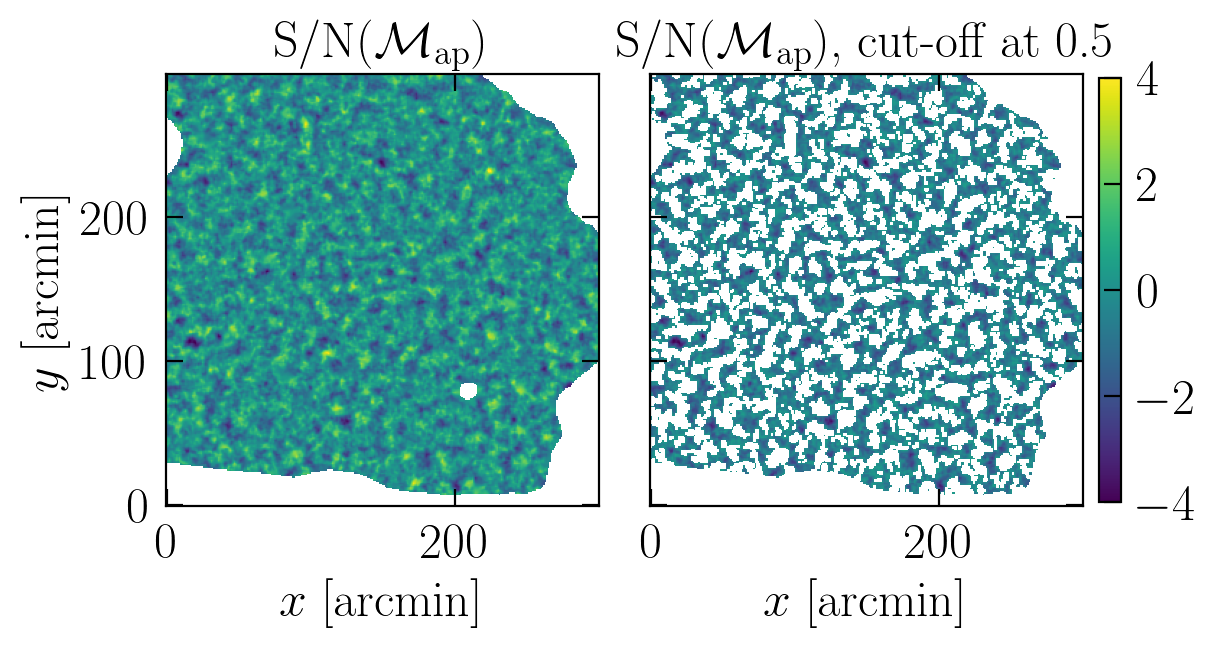}
    \caption{Example signal-to-noise map of aperture masses for a $5\times 5\,\mathrm{deg}^2$ sub-patch of one of the {\it Covariance Training Set} catalogue (left), and the same map when a threshold of 0.5 is applied (right). The white ``holes'' in the right map correspond to local maxima of the map and give rise to the topological ``features'' that are summarised in $\dgm_1$.}
    \label{fig:aperture_mass_map_with_cutoff}
\end{figure*}
\begin{figure}
    \resizebox{\hsize}{!}{\includegraphics{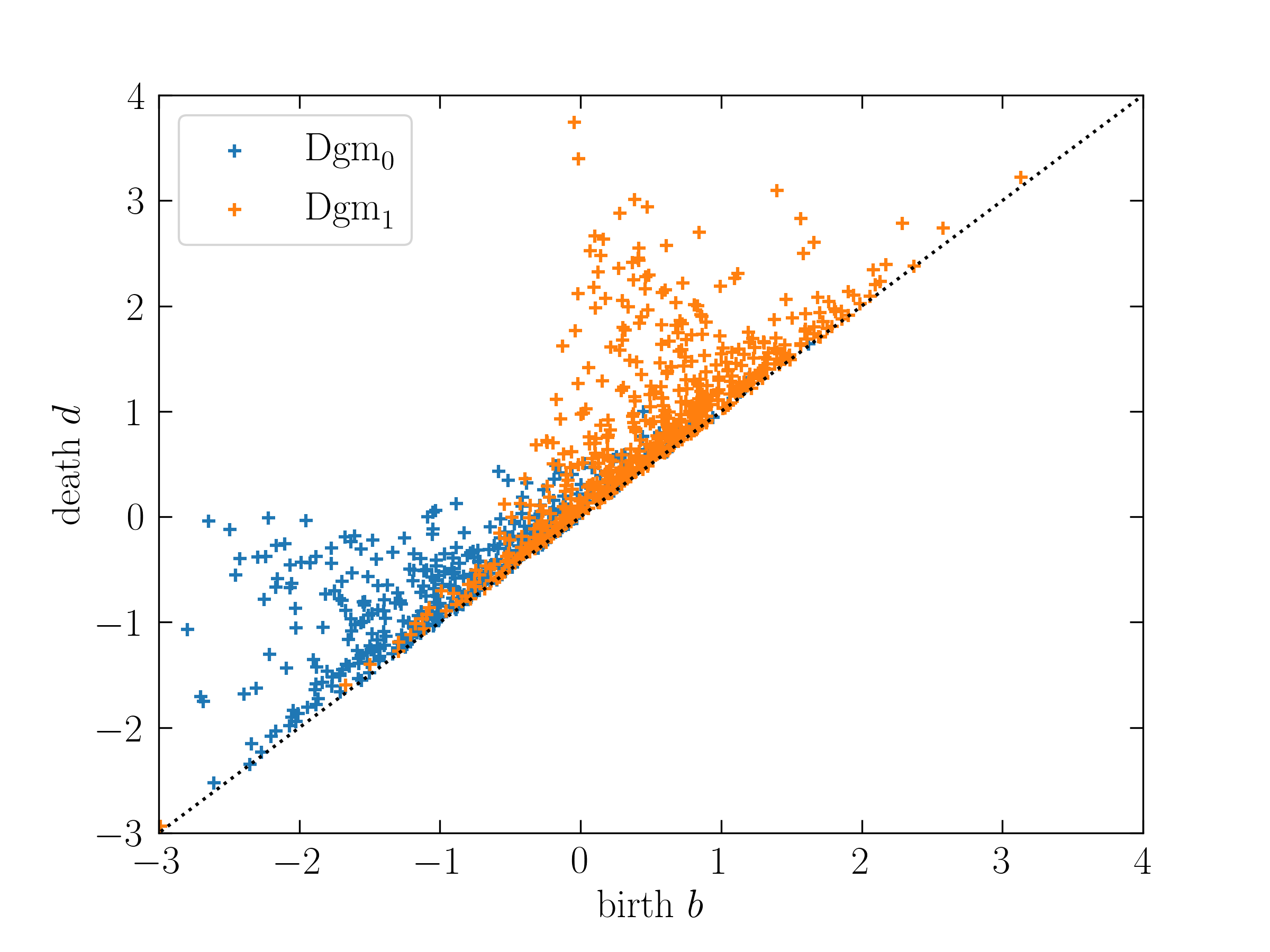}}
    \caption{The two persistence diagrams for the simulation shown in Fig.\ref{fig:aperture_mass_map_with_cutoff}. The blue crosses represent features of $\dgm_0$, the orange crosses represent features of $\dgm_1$. For visibility, only every 500-th feature is shown. Note that all points in this diagram lie above the diagonal.}
    \label{fig:persistence_diagram}
\end{figure}
\begin{figure*}
    \sidecaption
    \includegraphics[width=12cm]{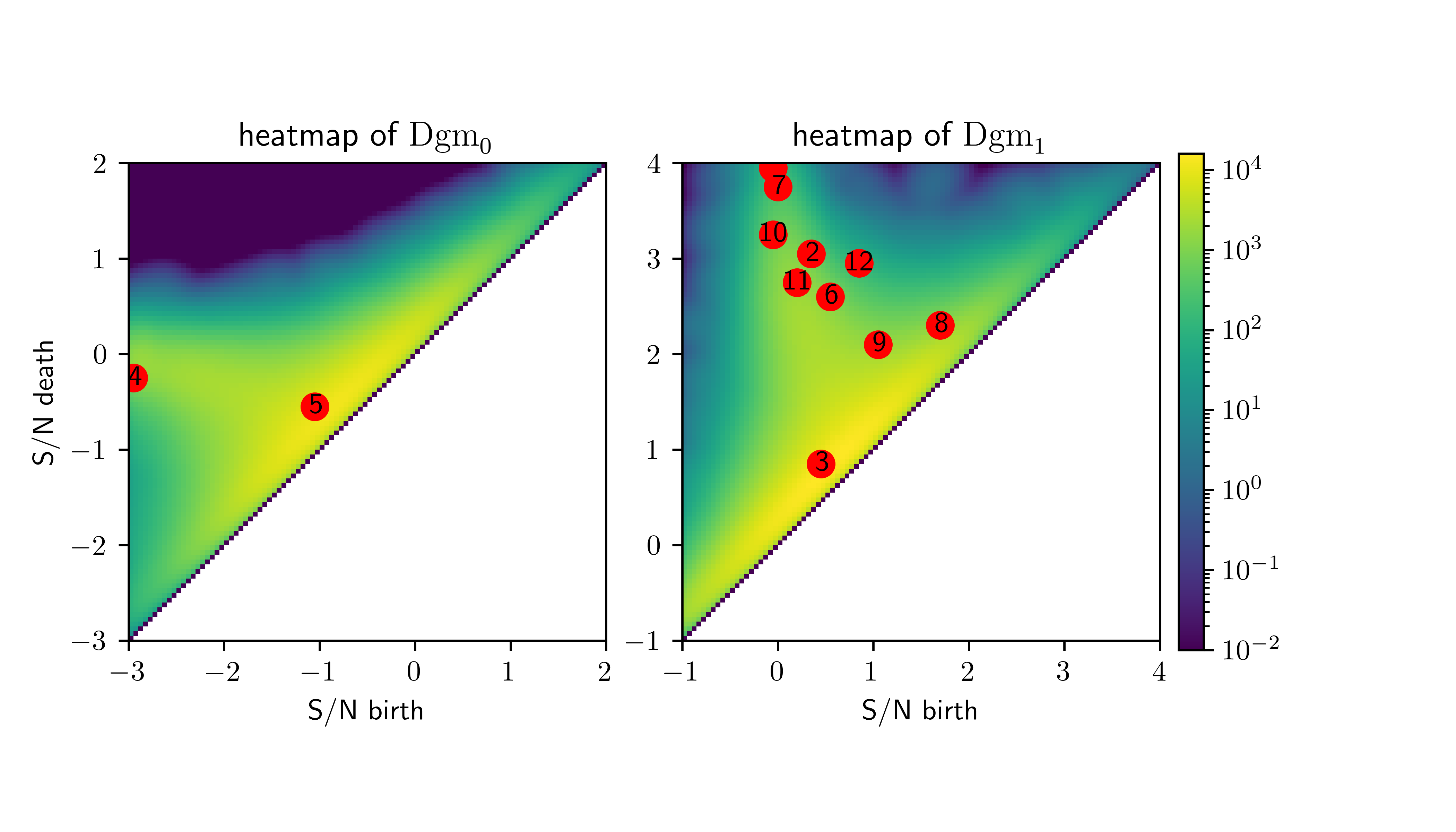}
    \caption{The heatmap of the persistence diagram in Fig. \ref{fig:persistence_diagram} with a scaling parameter of $t=0.2$ (for the computation of the Heatmaps, all features are taken into account, not just every 500-th as in Fig. \ref{fig:persistence_diagram}). The red points correspond to the evaluation points that were chosen by the $\chi^2$-maximiser outlined in Sect.~\ref{sec:data_compression}. The extracted data vector can be seen in Fig.~\ref{fig:data vector}.}
    \label{fig:heatmap}
\end{figure*}
\begin{figure}
    \resizebox{\hsize}{!}{\includegraphics[width=12cm]{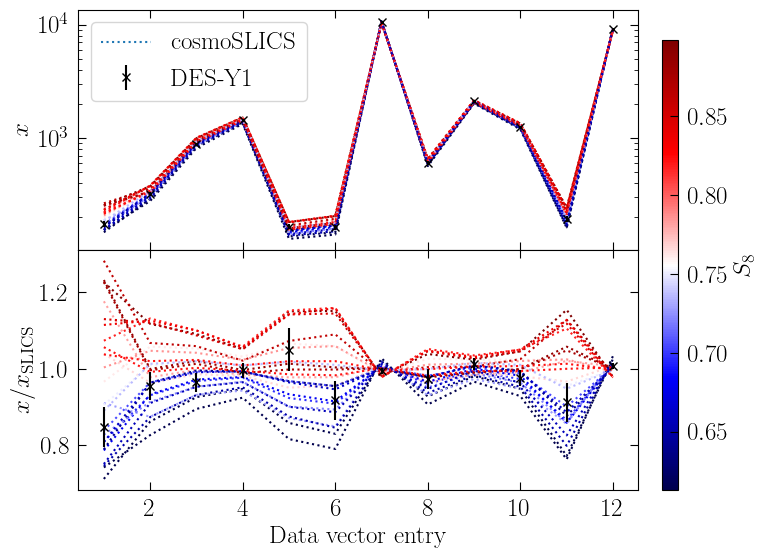}}
    \caption{The data vector for the individual cosmologies from our {\it Cosmology Training Set}, colour-coded by their respective value of $S_8$ (dotted lines) and the measured values in the DES-Y1 survey (black). For better visibility, the bottom panel shows the same data vector where all values are divided by the mean of our {\it Covariance Training Set}.}
    \label{fig:data vector}
\end{figure}
In this section, we give a short overview about the aspects of persistent homology that we use in the present paper. More detailed explanations can be found in Sect.~2.3 of H+21. For a general introduction to the topic that is geared towards its applications in data science, see \citet{CM:introductionTopologicalData} or \citet{OPT+:roadmapcomputationpersistent}; further information about the mathematical background can be found in \citet{Oud:Persistencetheory:quiver}.

Persistent homology is a technique from topological data analysis that allows to summarise the topological features of a sequence of spaces.
This is a versatile tool that can be applied in many different settings. However, the application that is relevant for the present article is that persistent homology gives a summary of the topological features of a map $f:X \to \mathbb{R}\cup \{\pm \infty \}$, where $X$ is in principle any (topological) space. (Here, the sequence of spaces is given by taking subsets of $X$ that consist of points where the value of $f$ lies below a certain threshold.)
In our setting, $X$ is a $(10\times 10\, \mathrm{deg}^2)$ tile of the sky (which we interpret as a subset of $\mbR^2$ and represent by a $(600 \times 600)$ pixel grid) and $f$ is the function that assigns to every point its S/N value as defined in \cref{sec_calculating_aperture_mass}. An example of this can be found in \cref{fig:aperture_mass_map_with_cutoff}.

The persistent homology of each such map $f$ can be summarised by two \emph{persistence diagrams} $\dgm_0 = \dgm_0(f)$ and $\dgm_1 = \dgm_1(f)$. Each of these persistence diagrams is a collection of intervals $[b,d)$, where $b,d \in \mbR \cup \{\pm \infty \}$. As each such interval is determined by the two values $b < d $, one can equivalently see a persistence diagram as a collection of points $(b,d)$ in $(\mbR \cup \{\pm \infty \})^2$ that lie above the diagonal.  We will call such a point $(b,d)$ a \emph{feature} of the persistence diagram; $b$ is commonly called the \emph{birth} and $d$ is its \emph{death}.
Roughly speaking, the points in $\dgm_0$ correspond to the local minima of the function $f$ whereas the points in $\dgm_1$ correspond to the local maxima. In both cases, the difference $d-b$ of such a feature $(b,d)$ is called its \emph{persistence} and describes how much the corresponding extremum protrudes from its surroundings. 

The actual computation of the $\dgm_0$ or $\dgm_1$ associated to an S/N map $f$ is done as follows: 
As explained in \cref{sec_calculating_aperture_mass}, the S/N maps we compute are defined on a $(600 \times 600)$ pixel grid, and a subset of these pixels are masked. We set the value of every masked pixel to be $-\infty$ and compute the persistence diagrams $\dgm_0$ and $\dgm_1$ associated to this map\footnote{We use the \texttt{Cubical Complexes} module of the public software {\sc gudhi} \citep{gudhi:CubicalComplex}.} (see \cref{fig:persistence_diagram}).
As explained in H+21 (Sect.~2.3.3), relative homology offers a natural way to work with persistent homology in the presence of masks.
The idea here is that a feature in $\dgm_i$ that is of the form $(-\infty, d)$ corresponds to a minimum or maximum of $f$ that originates from a masked area as these are the only points where $f$ takes the value $-\infty$.
%\footnote{Because of our setup here, were $f$ is a map from a contractible space $X$ to $\mbR\cup \{- \infty \}$, there are no features of the form $(b,\infty)$.} 
This is why we do not actually work with $\dgm_0$ or $\dgm_1$ but instead define `masked' persistence diagrams $\dgm_0^M$ and $\dgm_1^M$ as follows: The persistence diagram $\dgm_0^M$ is obtained from $\dgm_0$ by simply removing all features coming from the masks, i.e.~all points of the form $(-\infty,d)$. To get $\dgm_1^M$ from $\dgm_1$, we again start by removing all points of the form $(-\infty,d')$, but then for every feature of the form $(-\infty,d)$ in $\dgm_0$ (so those that got removed when transforming $\dgm_0$ into $\dgm^M_0$), we add a new feature  $(d,\infty)$ to  $\dgm_1^M$\footnote{$\dgm_0^M$ and $\dgm_1^M$ are the persistence diagrams associated to the persistence modules of the homology relative to the masked regions $M$. This is why $\dgm_1^M$ is not simply obtained by removing all mask features from $\dgm_1$. For more explanations, see H+21, Sect.~2.3.3.}.

\begin{comment}
In the end, we want to convey the message that the persistent homology is basically the same as the persistence diagram that one can obtain. So now it's all about studying these diagrams and statistics on them.
\end{comment}

\subsection{Persistence Statistics}
\label{sect:pers_stats}
From the calculations described in the previous section, we obtain for each S/N map two persistence diagrams  $\dgm_0^M$ and $\dgm_1^M$.
In order to carry out a statistical analysis of these persistence diagrams, one needs to be able to compute expected values and covariances. A priori, a persistence diagram is a particular collection of points in $(\mbR \cup \{\pm \infty \})^2$ and there is no canonical way of computing distances, sums and averages of such collections. There are different approaches to overcome these difficulties. Most of them proceed by converting persistence diagrams into elements of a suitable vector space and then using tools for statistics and data analysis in this space.
An overview of different options to perform statistics on persistence diagrams can be found in \cite{CM:introductionTopologicalData}, in particular Section 5.9 and in \cite{PXL:PersistentHomologybased}.
For this work, we tested three different approaches to the problem: persistent Betti numbers, persistence landscapes and heatmaps. All of these convert persistence diagrams into elements of certain function spaces.

\emph{Persistent Betti numbers} are the probably most direct approach and were used in H+21. They represent a persistence diagram $\dgm_i$ by the function $\beta_i : (\mbR \cup \{\pm \infty \})^2 \to \mbR$, where $\beta_i(x,y)$ is the number of points $(b,d)$ in $\dgm_i$ that lie to the ``upper left'' of $(x,y)$, i.e.~such that $x\leq b$ and $d \leq y$. 
%Note that these functions actually only take value in the natural numbers $\mbN$. Hence, taking for example the average of two such functions does not necessarily give you the persistent Betti number of another persistence diagram.
For more explanations about persistent Betti numbers, see H+21, Section 2.2.3 and Appendix B.

\emph{Persistent landscapes} are a more elaborate alternative, introduced in \cite{Bub:Statisticaltopologicaldata}  and already successfully used in applications, e.g.~in \citet{CSL+:Vectorizedpersistenthomology} and
\citet{KBNH:Usingpersistenthomology}.
However, we were not able to set them up in a way that led to competitive results. We suspect that the reason for this was the great number of features (around 500,000 per line-of-sight) in our persistence diagrams. The problem we were facing was that because of this number of features, we obtained a very large and noisy data vector. We reduced its dimension using a principal component analysis, similar to \citet{KBNH:Usingpersistenthomology}, but unfortunately, the quality of the resulting data was not good enough to obtain sufficiently tight bounds on the cosmological parameters. This might change when the principal component analysis can be applied to a less noisy data vector that is extracted from a larger training sample.

\emph{Heatmaps} are the method that worked best for us in the present setting. These are defined in the spirit of the multi-scale kernel introduced in \cite{RHBK:stablemultiscale}.
The idea is to replace each point in a persistence diagram by a Gaussian. More precisely, one considers the diagram as a discrete measure (i.e. a sum of Dirac delta distributions) on $D\subset \mbR^2$, where $D = \{(x,y) \in \mbR^2 | x < y \}$ and convolves this with a two-dimensional isotropic Gaussian distribution.
The result is for every value $t>0$ a continuous function $u_t(x,y): D \to \mbR$ that can be seen as a smoothed version of the persistence diagram. The value $t$ is called the \emph{scaling parameter} and determines how much smoothing is applied to the initial diagram. For an example of such a heatmap, see \cref{fig:heatmap}.

We compute the heatmaps in the following way: we first compute the persistence diagram $\dgm^M$ for the S/N maps of each tiled realisation of the DES-Y1 footprint, and for each tomographic bin. For each of these, we create a two-dimensional histogram of the persistence diagram with $100\times 100$ bins. For $\dgm_0$ our bins cover the S/N range $[-3,2]^2$, for $\dgm_1$ the bins cover the S/N range $[-1,4]^2$. The upper limit of 4 in the S/N maps avoids issues with source-lens coupling as elaborated in \citet{Martinet:2018}. All persistence features that lie outside of this range are projected to the edge of the respective bin ranges. %\textcolor{purple}{[JHD: `one tiled realisation of the DES-Y1 footprint' I guess you mean the full set of 19 tiles? I just need to make sure I use the same wording in the simulation section. Also, did we try a different binning scheme for the 2D histograms? 100 x 100 seems arbitrary, does it matter?]}\textcolor{OliveGreen}{[SH: Yes, that is what I mean. 100x100 is indeed arbitrary, but since we are smoothing with a kernel that is much larger than the bin size, I think it does not matter.]I   }

Afterwards, we convolve these histograms with a Gaussian kernel of scaling parameter $t=0.2$ using two-dimensional FFTs\footnote{We tried different values between $t=0.05$ and $t=0.4$. The results were stable with respect to these changes, and the value of 0.2 appears to be a good compromise between stability and precision.}.

\subsection{Data Compression}
\label{sec:data_compression}
To perform a Bayesian cosmological parameter inference, we compress the data provided by the persistence heatmaps. We explored several compression methods, which are discussed in App.~\ref{sec:app_data_compression}.

In the end we opted for an adaptation of our method developed in H+21; we iteratively build a data vector in the following way: As a first step, for each pixel $x$ of a heatmap we compute the mean squared difference between the single cosmologies of {\it cosmo}-SLICS and their mean, weighted by the inverse variance within the SLICS \begin{equation}
    \Delta x_\text{weighted} \equiv \sum_{i=0}^{25} \frac{\left(x_{\mathrm{cosmoSLICS},i}-\langle x_\mathrm{cosmoSLICS}\rangle\right)^2}{\sigma^2(x_\mathrm{SLICS})}\, .
    \label{eq:squared_difference_SLICS_cosmoslics}
\end{equation}
This $\Delta x_\text{weighted}$ describes the \emph{cosmological information content} of a pixel from the heatmap, as it quantifies how much its value varies between different cosmologies with respect to the expected standard deviation.
As the first point of our data vector we choose the one with the highest cosmological information content. Then we proceed to add more points in the following way: assuming we already have $n$ entries in our data vector, we determine the next entry  from the mean squared difference, weighted by the inverse sub-covariance matrix estimated from the SLICS. In other words: Let $\Delta \vec{x}_i \equiv \vec{x}_{\mathrm{cosmoSLICS},i}-\langle \vec{x}_\mathrm{cosmoSLICS}\rangle$ be the difference between the data vector of the $i$-th cosmology of {\it cosmo}-SLICS and the mean data vector of all {\it cosmo}-SLICS. For each pixel in the heatmap that is not already part of the data vector $\vec{x}$, we create a new data vector $\vec{x}'$ that contains this pixel, then we compute
\begin{equation}
    \hat{\chi}^2=\sum_{i=0}^{25} \Delta \vec{x}_i' C_\mathrm{SLICS}^{-1} \Delta \vec{x}_i'\; ,
\end{equation}
The pixel yielding the highest $\hat{\chi}^2$ is then added to the data vector and the procedure is repeated, until we have reached the desired amount of data points. Again, this serves to maximise the cosmological information content of our data vector with respect to the expected covariance. To ensure that our data vector follows a Gaussian distribution, we only consider elements of the heatmaps that count at least 100 features. We found that 12 data points per tomographic bin combination yields good results, but the dependence on the number of data points is weak. An example of such resulting data vector can be seen in Fig.~\ref{fig:data vector}.

While this method certainly does not capture the whole information residing in the heatmaps, this ``$\chi^2$-maximiser'' manages to capture most information and is therefore competitive with the other data compression methods.
A comparison is given in App.~\ref{sec:app_data_compression}.

\subsection{Two-point statistics}
The established methods to infer statistical properties of the matter and galaxy distribution concentrate on the second-order statistics such as the 2PCFs, their Fourier counterparts, the power spectra, or derived measures such as COSEBIs \citep{Schneider:2010,Asgari:2020}. The key advantage of these statistics over others is that, although they capture only the Gaussian information of the large-scale structure, they can be calculated analytically from the well-understood matter power spectrum $P(k,z)$. Indeed, the lensing power spectrum between galaxies of tomographic bin $i$ with redshift distribution $n^i(z)$ and those of tomographic bin $j$ with $n^j(z)$ is modelled in the Limber approximation as
\begin{equation}
    C_\ell^{ij} = \int_0^{\chi_\mathrm{H}} \frac{W^i(\chi)W^j(\chi)}{\chi^2}P\left(\frac{\ell+1/2}{\chi},z[\chi]\right) \dd \chi \,,
    \label{eq:C_ell}
\end{equation}
where $\chi_\mathrm{H}$ is the co-moving  distance to the horizon and $W(\chi)$ is the lensing efficiency defined as
\begin{equation}
    W(\chi) = \frac{3\Omega_\mathrm{m}H_0^2}{2c^2}\int_\chi^\infty \dd \chi'\, \frac{\chi(\chi'-\chi)}{\chi'a(\chi)}\,q(\chi') \;.
    \label{eq:lensing_efficiency}
\end{equation}
Here, $q(\chi)=n(z[\chi])\frac{\dd z[\chi]}{\dd \chi}$ is the line-of-sight probability density of the galaxies, $H_0$ the Hubble parameter and $c$ the speed of light. From the projected lensing power spectrum the cosmic shear correlation functions $\xi_\pm^{ij}$ are computed as
\begin{equation}
    \xi_\pm^{ij}(\vartheta) = \frac{1}{2\pi} \int_0^\infty C_\ell^{ij} J_{0,4}(\ell \vartheta) \,\ell \, \dd \ell
    \label{eq:xi_th}
\end{equation}
where $J_{0,4}$ are the Bessel functions of the first kind. To compute the theoretical two-point correlation functions we calculate the power spectrum $P(k)$ using the public {\sc Halofit} model \citep{Takahashi2012}.

We use the software \textsc{treecorr} \citep{Jarvis:Bernstein:2004} to  estimate the 2PCF $\Hat{\xi}_\pm^{ij}( \vartheta) $ from the simulations and the DES-Y1 lensing data, computed as
\begin{equation}
  \Hat{\xi}_\pm^{ij}( \vartheta) = \frac{  \sum_{a,b } w_a w_b\left[\epsilon^i_\mathrm{t}(\vec \theta_a)\epsilon^j_\mathrm{t}(\vec \theta_b) \pm \epsilon^i_\times(\vec \theta_a)\epsilon^j_\times(\vec \theta_b)\right]}{\sum_{a,b} w_a w_b S_a S_b}  \, , 
  \label{eq:xi_estimator}
\end{equation}
where the sums are over all galaxy pairs $(a,b)$ in tomographic bins $(i,j)$ that are inside the corresponding $\vartheta$-bin.
As in HD+21, we used 32 logarithmically spaced $\vartheta$-bins in the range
$[0.5\arcmin,475\arcmin\! .5]$, although not all angular scales are used in the parameter estimation (see the following section).

\subsection{Cosmological parameter estimation}
\label{sec:cosmological_parameter_estimation}
\begin{figure}
\resizebox{\hsize}{!}{\includegraphics{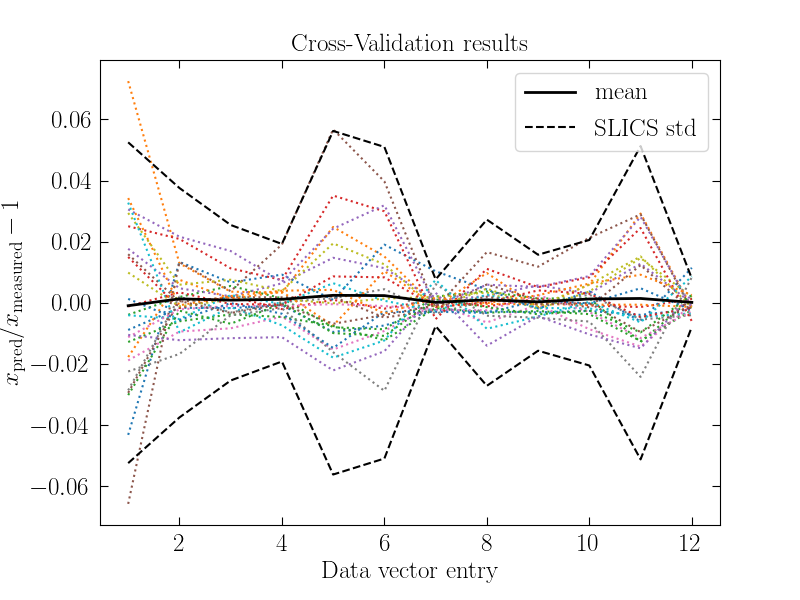}}
    \caption{The accuracy of the GPR emulator evaluated by a leave-one-out cross-validation, %for the part of the data vector corresponding to the combination of 
    shown here for the case where the aperture mass maps are constructed from the concatenation of
    all four tomographic redshift bins (i.e. no tomography). The $x$-axis depicts the data vector entry, the $y$-axis the relative difference between predicted and measured value. The 26 individual dotted lines correspond to one cosmology that is left out of the training set and then predicted, the solid black line is the mean of all dotted lines. The black dashed lines depict the standard deviation from the \emph{Covariance Training Set}. }
    \label{fig:cv_accuracy}
\end{figure}

As in H+21, we train a GPR emulator using data extracted from the 26 different {\it cosmo}-SLICS  models to interpolate our data vector at arbitrary cosmological parameters within the training range. We refer the reader to H+21 and HD+21 for more details on the emulator. 
We assess its accuracy by performing a leave-one-out cross-validation: we remove one cosmology of the {\it cosmo}-SLICS from our training sample and let the GPR-emulator predict this cosmology, training on the other 25. We repeat this procedure for all 26 cosmologies, and use the mean squared difference between predictions and truth as an estimate for the error of the emulator, which is typically well below the statistical error (compare Fig.~\ref{fig:cv_accuracy}). We then add this to the diagonal of our sample covariance matrix to account for uncertainties in the modelling.

An alternative method to estimate the uncertainty on the predictions is to use the error provided by the GPR emulator itself. We tested this method as well and found that, while this method is a bit slower (since the inverse covariance matrix needs to be re-computed in every step of the MCMC), it provides comparable, albeit slightly tighter constraints than the first method. In the end, we opted for the more conservative choice of estimating the modelling uncertainties via cross-validation.

As in HD+21, we then integrate our GPR emulator into the {\sc cosmoSIS} analysis pipeline \citep{arXiv:1409.3409} and infer the cosmological parameters by sampling the likelihood using the polychord sampling method \citep{Handley:2015}, which constitutes a good compromise between speed and accuracy \citep{Lemos:2022}. A few relatively minor changes to the {\sc cosmoSIS} likelihood module allow for an easy and fast joint analysis of both persistent homology statistics and shear two-point correlation functions.

\begin{figure}
    \centering
    \resizebox{\hsize}{!}{\includegraphics{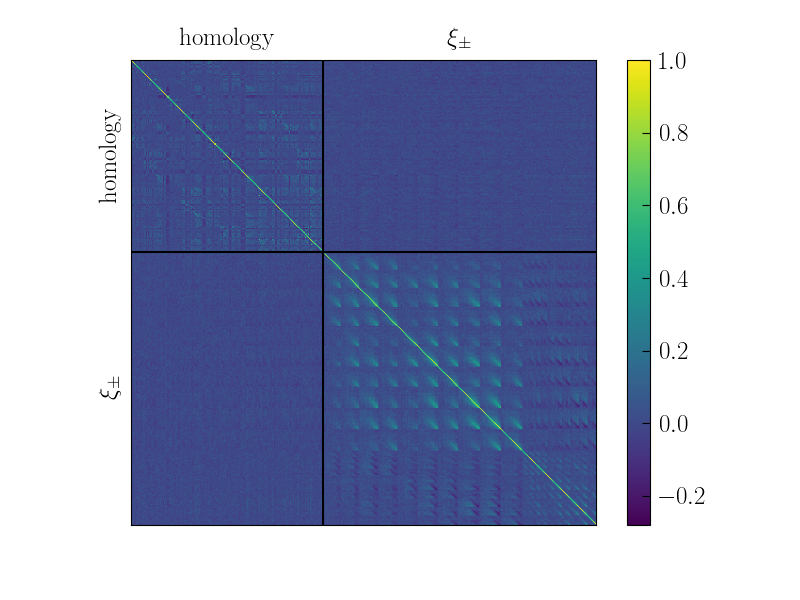}}
    \caption{The correlation matrix for a joint analysis with both persistent homology and two-point correlation functions.}
    \label{fig:covariance_matrix}
\end{figure}
\begin{table}
\caption{Prior ranges of cosmological and nuisance parameters in the likelihood analysis Priors on cosmological parameters are provided by the range of our {\it Cosmology Training Set}, prior on $\Delta m_i$ and on $A_{\rm IA}$  are from T+18, while those on $\dd z_i$ are from \citet{joudaki2020}.}
\label{tab:priors}
\centering
\begin{tabular}{c c c}
\hline\hline
Parameter & Prior Type & Prior range \\
\hline
$\Omm$ & uniform & $[0.1,0.55]$ \\
$h$ & uniform & $[0.6,0.82]$ \\
$\Omega_\mathrm{b}$ & delta & 0.0447 \\
$\tau$ & delta & 0.08 \\
$n_s$ & delta & 0.969 \\
$\sigma_8$ & uniform & $[0.53,1.3]$ \\
$w_0$ & uniform & $[-2.0,-0.5]$ \\
$w_a$ & delta & 0 \\
$S_8$ & uniform & $[0.6,0.9]$ \\
$A_\mathrm{IA}$ & uniform & $[-5,5]$ \\
baryon feedback & uniform & $[0,2]$ \\
$\Delta m_1$ & Gaussian & $\mu=0.12,\;\sigma=0.023$ \\
$\Delta m_2$ & Gaussian & $\mu=0.12,\;\sigma=0.023$ \\
$\Delta m_3$ & Gaussian & $\mu=0.12,\;\sigma=0.023$ \\
$\Delta m_4$ & Gaussian & $\mu=0.12,\;\sigma=0.023$ \\
$\dd z_1$ & Gaussian & $\mu=0,\;\sigma=0.008$ \\
$\dd z_2$ & Gaussian & $\mu=0,\;\sigma=0.014$ \\
$\dd z_3$ & Gaussian & $\mu=0,\;\sigma=0.011$ \\
$\dd z_4$ & Gaussian & $\mu=0,\;\sigma=0.009$ \\

\end{tabular}
\end{table}

Finally, we estimate our sample covariance matrix from the 124 survey realisations of the {\it Covariance Training Set}. Specifically, we compute a matrix for each of the 10 different realisations of the shape noise, and use the average over those 10 covariance matrices as our best estimate. Here, we randomly distribute the 124 lines-of-sight for the 19 regions to avoid over-estimating the sample variance (compare HD+21). Further, since the inverse of a simulation-based covariance matrix is generally biased \citep{Hartlap:2007}, we mitigate this effect by adopting a multivariate $t$-distribution likelihood \citep{Sellentin:2016}. The extracted covariance matrix can be seen in Fig.~\ref{fig:covariance_matrix}, the priors used for cosmological parameter estimation are listed in Tab.~\ref{tab:priors}.

\section{Mitigating systematic effects}
\label{sec:systematics}
Our cosmological parameter analysis needs to account for systematic effects that are known to affect cosmic shear data. The most important ones for this work are intrinsic alignments of source galaxies,  baryonic physics, multiplicative shear bias and uncertainties in the redshift estimation of galaxies \citep{Mandelbaum18}. On top of these, limits in the force resolution of the {\it cosmo}-SLICS might introduce a bias into our modelling, plus source clustering can produce systematic differences between the data and the simulations, in which the latter is absent. While we investigate the former, the latter has been shown in HD+21 to be largely subdominant in the aperture mass statistics measured in the DES-Y1 data and is therefore neglected here.

In this section, we explain how these systematic effects affect our 2PCFs and persistent homology measurements, and detail the mitigation strategies we chose to account for their impact.

\subsection{2PCF}

We use the public modules in {\sc cosmoSIS} to marginalise over the impact of intrinsic alignments. Following T+18, we model IA with the non-linear alignment model \citep[][hereafter referred to as the NLA model]{BridleKing}, which adds a contribution to the matter power spectrum that propagates into the lensing signal following Eqs. (\ref{eq:C_ell}) and (\ref{eq:xi_th}). More sophisticated IA models including tidal torque terms \citep[notably the Tidal Alignment and Tidal Torque  model, or TATT][]{Blazek2019} have been used recently in cosmic shear analyses but there is no clear evidence that the data prefer such a model over the NLA \citep[T+18,][]{DESY3_Secco}. The NLA model can have multiple parameters (amplitude, redshift dependence, luminosity dependence, pivot scales, colour), however we follow HD+21 and vary only the amplitude ($A_{\rm IA}$) and luminosity ($\alpha$) parameters, considering no other dependencies. This is justified by the weak constraints that exist on them in the DES-Y1 data (compare T+18). The parameters $A_{\rm IA}$ and $\alpha$ are allowed to vary in the range [-5.0, 5.0].

Following the fiducial DES-Y1 choices, the impact of baryon feedback is minimised by cutting out angular scales in the $\xi_\pm$ statistics where unmodelled baryonic physics with a strong AGN model\footnote{The power spectrum of the OWLS AGN model \citep{OWLS} is used for this assessment.} could shift the data by more than 2\%. We therefore exclude from our analysis the same small scales as those of T+18, which are different for $\xi_+$ and $\xi_-$, and further vary with redshift.
  
The shear inference is obtained with the {\sc Metacalibration} method in this work, which has a small uncertainty that can be captured by a shape calibration factor $\Delta m$, which multiplies the observed ellipticities as $\epsilon_{1/2} \rightarrow \epsilon_{1/2}(1+\Delta m)$. As described in T+18, $\Delta m$ is a nuisance parameter that we sample by a Gaussian distribution with a width of 0.023,  centred on 0.012 when analysing the data, and on zero when analysing simulations. {\sc cosmoSIS} includes this nuisance on the  two-point function model directly, namely $\xi_{\pm}^{ij} \rightarrow \xi_{\pm}^{ij}(1+\Delta m^i)(1+\Delta m^j)$. The priors on $m$ are listed in Tab.~\ref{tab:priors}.
  
Photometric errors in the 2PCFs are mitigated by using the generic module within {\sc cosmoSIS}, which shifts the $n^i(z)$ by small bias parameters $\Delta z^i$ and updates accordingly the lensing predictions. These bias parameters are sampled from Gaussian distributions with widths corresponding to the posterior DIR estimates of the mean redshift per tomographic bin `$i$', also tabulated in Tab.~\ref{tab:priors}.
  
\subsection{Persistent homology}

\begin{figure*}
    \sidecaption
    \includegraphics[width=12cm]{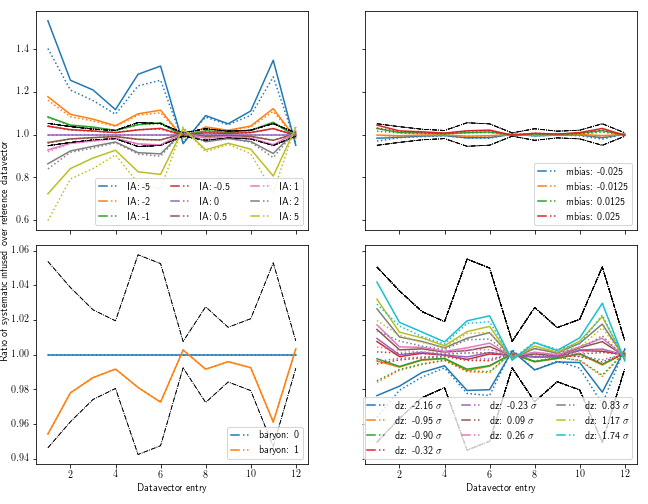}
    \caption{The impact of the main systematic effects on the data vector. For each systematic, we show the measured (solid line) and interpolated (dotted line) ratio of the systematic-infused data vector over a reference data vector. For simplicity, we only show the results for the combination of all four tomographic redshift bins. The black dashed lines correspond to the $1\sigma$ standard deviation estimated from the {\it Covariance Training Set}.}
    \label{fig:systematic_effects}
\end{figure*}

As mentioned in Sec. \ref{subsec:mocks}, we assess the impact of systematics on the topology of aperture mass maps by using the {\it Systematics Training Set}, which are numerical simulations specifically tailored for this exercise. Following HD+21, we neglect the cosmology scaling of these systematics and only evaluate their relative impact at the fiducial cosmology. We find that the overall impact of systematic effects is sufficiently well captured by a linear modelling strategy: for each systematic effect with respective nuisance parameter $\lambda$ (i.e.~$A_{\rm IA}$ for intrinsic alignments, $\Delta z$ for redshift uncertainties, $\Delta m$ for multiplicative bias and $b_{\rm bar}$ for baryons), we measure the impact $\vec{x}_{\rm sys}(\lambda )$ on the measured data vector from the associated {\it Systematics Training Set} and fit each point of the data vector with a straight line:
\begin{equation}
    \vec{x}_{\rm sys}(\lambda) = \vec{m_x}\lambda+\vec{x}_{\mathrm{nosys}} \; .
\end{equation}
In particular, $\vec{x}(0)\equiv \vec{x}_\mathrm{nosys}$ is the data vector which is not impacted by any systematic effects. For a given set of values of the nuisance parameters, we combine these different sources of uncertainty to model our systematics-infused data vector as:
\begin{equation}
    \vec{x}_\mathrm{sys} = \vec{x}_\mathrm{nosys} + \vec{m}_\mathrm{IA}A_{\rm IA} + \vec{m}_\mathrm{bar}b_{\rm bar} + \vec{m}_\mathrm{\dd z}\Delta z + \vec{m}_\mathrm{\Delta m}\Delta m \; .
\end{equation}
While this certainly constitutes a simplified approach that does not capture potential cross-correlations between different systematic effects nor any cosmology dependence, we consider it sufficient at the current level of uncertainties (compare Fig.~\ref{fig:systematic_effects}).

To compute $\vec{m}_\mathrm{IA}A_{\rm IA}$, the Intrinsic Alignments mocks are infused with $A_{\rm IA}$ values of [-5.0, -2.0, -1.0, 0.5, 0.0, 0.5, 1.0, 2.0 and 5.0], however we set the redshift dependence to zero, given the weakness of the constraints on this parameter in the DES-Y1 data (see T+18). We report in the upper left panel of Fig.~\ref{fig:systematic_effects} the fractional effect on the signal, and observe that positive IA suppresses the elements of the data vector. The is caused by the partial cancellation of the lensing signal by IA, which attenuates the contrasts in the aperture mass maps, which translates in a topological structure that  has less features. The figure also presents the results as modelled by the linear interpolation, which reproduces the nodes on which the training was performed to a sufficient accuracy, indicating that our approach is adequate to model IA, at least for the range of $A_{\rm IA}$ values tested here.

The impact of shear calibration uncertainty is modelled by measuring the statistics for ellipticities modified with four values of $\Delta m$, namely -0.025, -0.0125, 0.0125, and 0.025, and once again fitting a straight line through each element of the homology data vector as a function of $\Delta m$. The results are presented in the upper right panel of  Fig.~\ref{fig:systematic_effects}, showing that within this range, shear calibration affects the statistics by less than one per cent except for two elements, which are affected by up to 4\%.

Photometric uncertainties are modelled from the dedicated {\it Systematics Training Set} in which the $n(z)$ in each tomographic bin has been shifted by 10 values $\Delta z^i$, from which we are once again able to fit a linear response for each element of the data vector. In the case of cross-redshifts, the mean of all shifts is used to compute the derivative, as in HD+21. The lower right panel of Fig.~\ref{fig:systematic_effects} shows the impact on the data vector, which is sub-dominant compared to the IA. This is largely due to the tight priors on  $\Delta z^i$ that we are able to achieve with the DIR method, as derived in \citet{joudaki2020} and reported in Tab.~\ref{tab:priors}.

The {\it Magneticum} simulations are used in a similar way to test the impact of baryonic feedback, with the main difference that we can only fit $\vec{m}_{\rm bar}$ on two points: the simulations with and without baryons ($b_{\rm bar}$ = 1 and 0, respectively). We nevertheless apply the same methodology here, which allows us to interpolate between these two cases to mimic milder models (e.g. $b_{\rm bar}$ = 0.5), and even to extrapolate and explore stronger feedback models ($b_{\rm bar}$ >1.0). The lower right panel of Fig.~\ref{fig:systematic_effects} shows that baryonic feedback with $b_{\rm bar}$ = 1.0 has almost as much importance as a IA model with $A_{\rm IA}$= 1.0 and should therefore not be neglected in this analysis.

The last systematic effect that we include in this analysis is the impact of the force accuracy in the $N$-body simulations that are used in the modelling. We inspect the difference between the data vector measured from the high-accuracy mocks to that of the main {\it Cosmology Training Sample} at the fiducial cosmology and find that the overall impact of this effect is sub-dominant to the sample variance of the SLICS. Nevertheless, we measure the ratio of the high-resolution data vector over the fiducial one and apply it as a correction factor to re-calibrate our model in the analysis of observed data.

\subsection{Mitigation strategy}
\begin{figure*}
    \sidecaption
    \includegraphics[width=6cm]{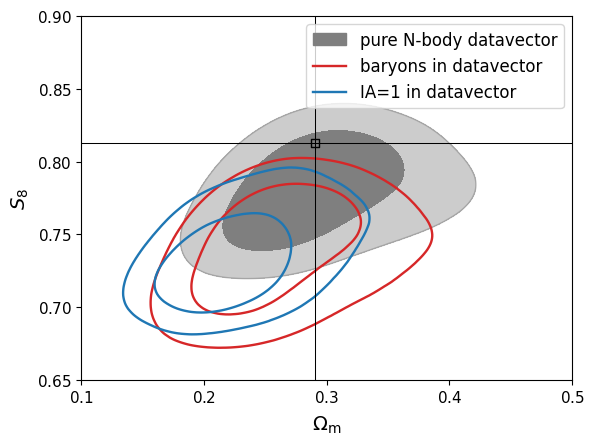}
    \includegraphics[width=6cm]{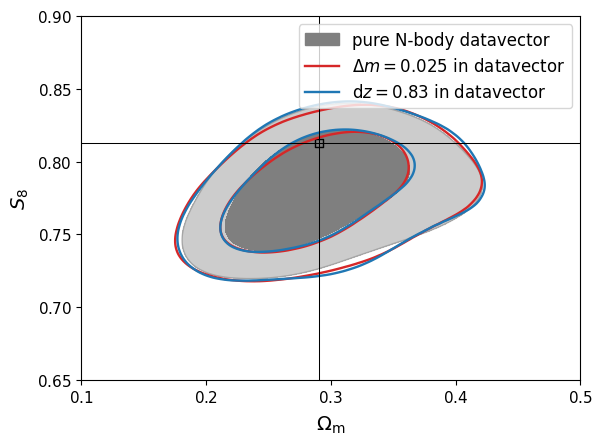}
    \caption{The impact of unmodelled systematic biases on the posterior of a likelihood analysis with heatmaps. In all cases, we do not marginalise over any systematic effects. The target data vector is then infused with one systematic bias, and we run a likelihood analysis for this infused data vector. For comparison we show the constraints on a data vector that is not infused by systematics (grey). Note that the values of the $\dd z$ shifts are given in units of the standard deviation of the $\dd z$ prior (compare Tab.~\ref{tab:priors}).}
    \label{fig:bias_systematics}
\end{figure*}
\begin{figure*}
\sidecaption
\includegraphics[width=6cm]{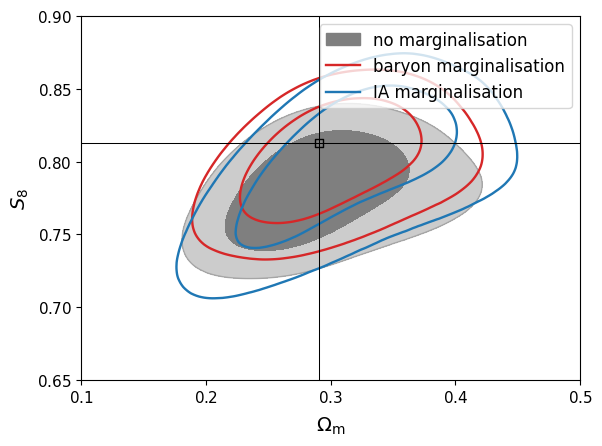}
\includegraphics[width=6cm]{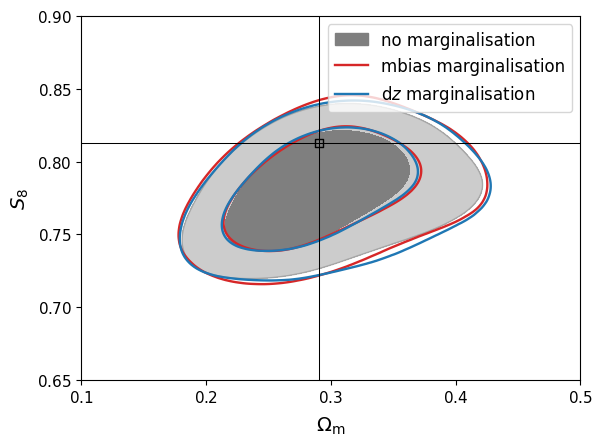}
\caption{The effects of marginalising over different systematic effects. In all cases we perform a likelihood analysis on mock data, marginalising over one systematic effect. For comparison we show the constraints we achieve when we do not marginalise over any systematics (grey). The case where we marginalise over all systematics corresponds to the blue contours in Fig.~\ref{fig:mcmc_joint}}.
\label{fig:marginalisation_systematics}
\end{figure*}

We further estimate the impact of the different systematic effects on the cosmology inference by running likelihood analyses for data vectors that have been infused with one systematic effect, while keeping these unmodelled. The results of these tests can be seen in Fig.~\ref{fig:bias_systematics}. We observe that the baryons have a small impact on the inferred $\Omega_{\rm m}$, and can bias $S_8$ by 1$\sigma$, assuming $b_{\rm bar}$=1.0. Unmodelled IA (with $A_{\rm IA}$=1.0) tend to bias both $\Omega_{\rm m}$ and $S_8$ towards lower values; both photometric redshift uncertainties and multiplicative shear bias have a minor impact on the posterior constraints, given the tight priors available on $\Delta z$ and $\Delta m$.

We finally investigate how marginalisation over the different systematic biases changes the posterior contours in our likelihood analysis in Fig.~\ref{fig:marginalisation_systematics}. 
We find that marginalisation over baryonic effects and intrinsic alignments both decrease the constraining power on $\Omm$ and $S_8$ by about $25\%$, whereas the marginalisation over multiplicative shear biases and photometric redshift uncertainties have a negligible impact.
%Specifically, we compare the case where a full marginalisation  is included (green contour) to the `no-systematics' case (dark blue contour) and notice that the former shifts towards larger values of $\Omega_{\rm m}$, \textcolor{purple}{[JHD: why? It should be centred, and consistent with the truth? Could that be caused by the IA mocks? Can you try to marginalise over all but IA perhaps?]}. Marginalising only over photometric error or shape calibration has a negligible impact on the contours \textcolor{purple}{[again why? shouldn't the contour puff up a little?]}. We also check we recover unbiased results when analysing baryon-infused data but marginalising that effect (magenta lines). The same exercise applied to the IA mocks results in ...\textcolor{purple}{[ why is there a shift towards high $\Omega_{\rm m}$?}

Both analyses suggest that the impact of systematic effects on persistent homology statistics is noticeable, but not severe, and that our marginalisation strategies work as expected.

\section{Validation}
\label{sect:validation}

In this work, we want to investigate whether a likelihood analysis of tomographic cosmic shear data with persistent homology is feasible, and whether a joint analysis with two-point statistics yields more information than an analysis that solely utilises two-point statistics. For this purpose, we perform three likelihood analyses of the same mock data extracted from the {\it Covariance Training Set}: one solely with two-point correlation functions that we model within the {\sc cosmoSIS} pipeline, one solely with our persistent homology method, and finally the combined analysis.
\begin{figure*}
    \centering
    \includegraphics[width=17cm]{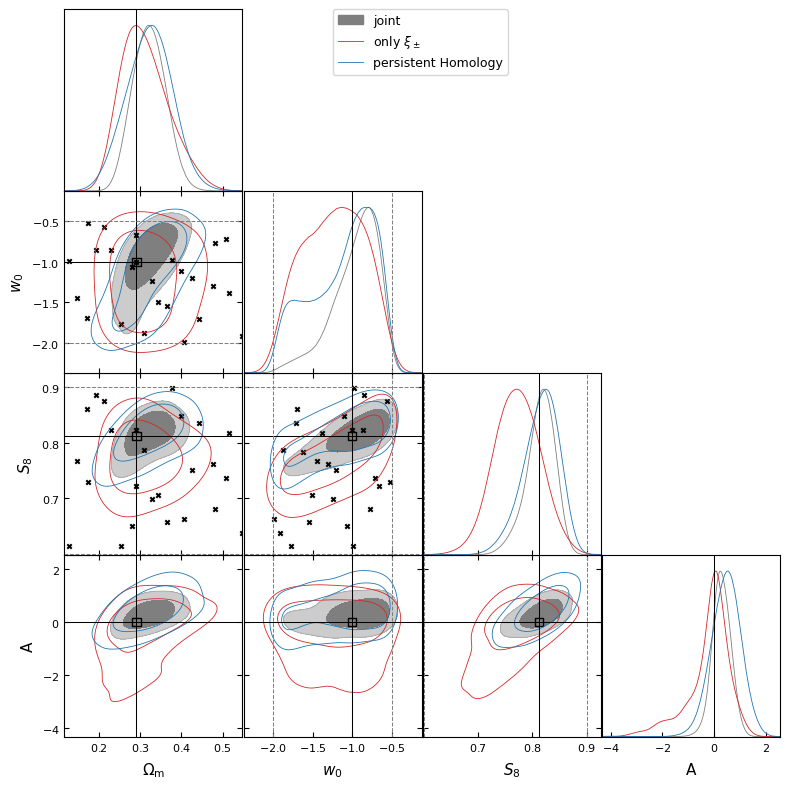}
    \caption{The results of likelihood analyses for DES-Y1 mock data. We show the results for two-point statistics (red),  persistent homology (blue), and for the  joint analysis (grey, filled). The dotted lines show the prior ranges, the solid black lines visualise the true value of each parameter, and the black crosses denote the nodes of our {\it Cosmology Training Set}. The complete results can be seen in Fig.~\ref{fig:mcmc_joint_full}, the marginalised posterior constraints can be seen in Tab.~\ref{tab:parameter_constraints}.}
    \label{fig:mcmc_joint}
\end{figure*}

As can be seen in Fig.~\ref{fig:mcmc_joint} and Tab.~\ref{tab:parameter_constraints}, the persistent homology analysis is already able to constrain $S_8$ better than the two-point analysis ($S_8=	0.817^{+0.040}_{-0.028}$ for persistent homology versus $S_8=0.772\pm 0.043$ for two-point statistics). However, a joint analysis offers several additional benefits. While two-point statistics are able to constrain the parameter to $A_\mathrm{IA}=-0.19^{+0.90}_{-0.40}$, persistent homology yields $A=0.47^{+0.64}_{-0.56}$ and a joint analysis is able to reduce the error bars to $A_\mathrm{IA}=0.29\pm 0.36$. Apart from tighter constraints on $S_8$ ($S_8 = 0.815^{+0.030}_{-0.021}$ for a joint analysis), a joint analysis also yields competitive lower limits on the equation-of-state parameter of dark energy ($w_0>-1.14$ at $68\%$ confidence), while two-point statistics are unable to place any constraints on this parameter, with our choice of sampling method.

Most importantly, all cosmological and nuisance parameters are recovered within $1\sigma$. We thus conclude that our analysis pipeline has been validated and move on towards a cosmological parameter analysis of real data.

\section{Results}
\label{sect:results}
\begin{figure*}
    \centering
    \includegraphics[width=17cm]{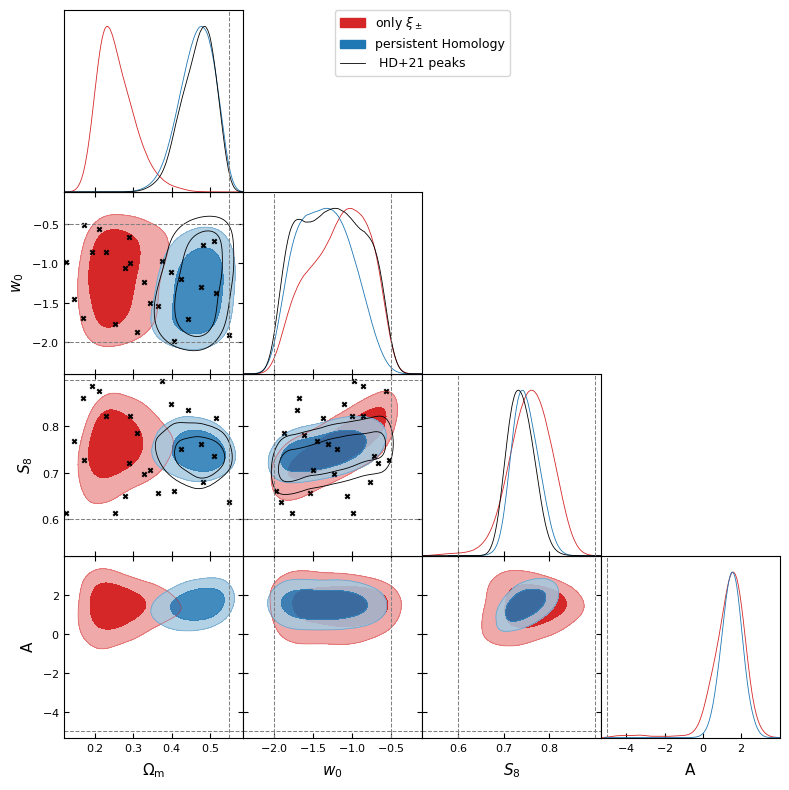}
    \caption{The results of our likelihood analyses for the DES-Y1 survey. We show the results for two-point statistics (red),  persistent homology (blue), and report as well the constraints achieved by HD+21 with peak count statistics (black). The dotted lines denote the prior ranges, the black crosses denote the nodes of our {\it Cosmology Training Set}. The complete results including nuisance parameters can be seen in Fig.~\ref{fig:mcmc_joint_des_full}, while the marginalised posterior constraints can be seen in Tab.~\ref{tab:parameter_constraints}.}
    \label{fig:mcmc_joint_des}
\end{figure*}

\begin{table*}
\sidecaption
\caption{Posterior $68\%$ confidence intervals on cosmological and nuisance parameters from the likelihood analyses in Fig.~\ref{fig:mcmc_joint} and Fig.~\ref{fig:mcmc_joint_des}. Constraints are only cited if the value of the marginalised posterior does not surpass $13.5\%$ at the edge of the priors \citep{asgari2021}.}
\label{tab:parameter_constraints}
\centering
\begin{tabular}{c c c c c}
\hline\hline
Method & $\Omega_\mathrm{m}$ & $S_8$ & $w_0$ & $A$ \\[0.2em]
\textbf{Validation (mock data)} & & & & \\
\hline
 & & & & \\[-0.8em]
persistent homology & $ 0.323^{+0.059}_{-0.053}$ & $ 0.817^{+0.040}_{-0.028}$ & $-$ & $ 0.47^{+0.64}_{-0.56}$ \\[0.1em]
$\xi_\pm$ & $ 0.311^{+0.046}_{-0.069}$ & $ 0.772\pm 0.043$ & $-$ & $ -0.19^{+0.90}_{-0.40}$ \\[0.1em]
joint & $ 0.321\pm 0.040$ & $ 0.815^{+0.030}_{-0.021}$ & $ >-1.14$ & $ 0.29\pm 0.36$ \\[0.2em]
\textbf{DES-Y1 data} & & & & \\
\hline
 & & & & \\[-0.8em]
persistent homology & $ 0.468^{+0.051}_{-0.036}$ & $ 0.747^{+0.025}_{-0.031}$ & $ <-1.04$ & $ 1.54\pm 0.52$ \\[0.1em]
$\xi_\pm$ & $ 0.256^{+0.034}_{-0.058}$ & $ 0.759^{+0.049}_{-0.042}$ & $ >-1.47$ & $ 1.33^{+0.92}_{-0.56}$ \\
\end{tabular}
\end{table*}
Having validated our analysis pipeline, we now use it to perform our cosmological parameter analyses using on the DES-Y1 data. In order to do that, we split the source galaxy catalogue into the same 19 tiles as our mock data and compute the persistence statistics as well as the two-point correlation functions for each tile individually.

The results can be seen in Fig.~\ref{fig:mcmc_joint_des}. We observe that neither persistent homology nor two-point correlation functions are able to place meaningful constraints on the equation of state parameter for dark energy, $w_0$. For the matter clustering parameter $S_8$, the constraints from persistent homology ($S_8=0.747^{+0.025}_{-0.031}$) are tighter than, but fully consistent with, the constraints from two-point correlation functions ($S_8=0.759^{+0.049}_{-0.042}$). The same goes for the amplitude of galaxy intrinsic alignments ($A=1.54\pm 0.52$ for persistent homology and $A=1.33^{+0.92}_{-0.56}$ for two-point correlation functions). In particular this implies that persistent homology rules out the case of no intrinsic alignments roughly at the $3\sigma$ level. Interestingly, the constraints for the matter density parameters are not consistent ($\Omm = 0.468^{+0.051}_{-0.036}$ for persistent homology and $\Omm = 0.256^{+0.034}_{-0.058}$ for two-point correlation functions). \citet{Hamana:2020} observed a similar trend when observing data from the HSC: While a real- and Fourier-space analysis yield perfectly consistent values for $S_8$, a slight tension between the $\Omm$ constraints can be observed in their Fig.~15. We observe a much larger tension that prevents us from performing a joint parameter analysis, which would tighten the $S_8$-constraints considerably. We discuss this in more detail in App.~\ref{sec:app_tension}, where we show that such a tension arises in about $0.5\%$ of all cases due to a mere statistical fluctuation.

Comparing our results with the ones from peak count statistics, where HD+21 measured $S_8=0.737^{+0.027}_{-0.031}$ on the same data set, we observe remarkably consistent results (compare Fig.~\ref{fig:mcmc_joint_des}). The trend towards high values of $\Omm$ can also be observed in HD+21 (see in particular Fig.~17 and 18 in HD+21). This is particularly interesting since their constraints have been achieved using a pipeline that is fully independent from ours, utilising a different statistic on signal-to-noise maps of aperture masses constructed with an independent code, albeit based on the same set of $N$-body simulations. Furthermore, we can see that the constraints achieved from persistent homology out-perform the ones from peak statistics, albeit not by much. This improvement is still significant, since contrary to HD+21 we include an error estimate for the emulator and a marginalisation over intrinsic alignments and baryonic effects, which decrease the constraining power of our analysis pipeline.

Comparing our results to T+18, we see that our results from two-point statistics are a bit different ($S_8=0.777^{+0.036}_{-0.038}$ in T+18 and $S_8=0.759^{+0.049}_{-0.042}$ here), which is driven mainly by the different redshift distribution estimates \citep[as shown in][]{joudaki2020}. Considering the intrinsic alignment effect, we achieve consistent, but tighter constraints (compare the NLA case of Fig.~16 in T+18). Regarding the tension we measure for $\Omm$, T+18 report $\Omm=0.274^{+0.073}_{-0.042}$, which is fully consistent with our results from two-point correlation functions and also disagrees with our constraints from persistent homology.

\section{Discussion}
\label{sect:discussion}
\begin{figure}
    \centering
    \resizebox{\hsize}{!}{\includegraphics{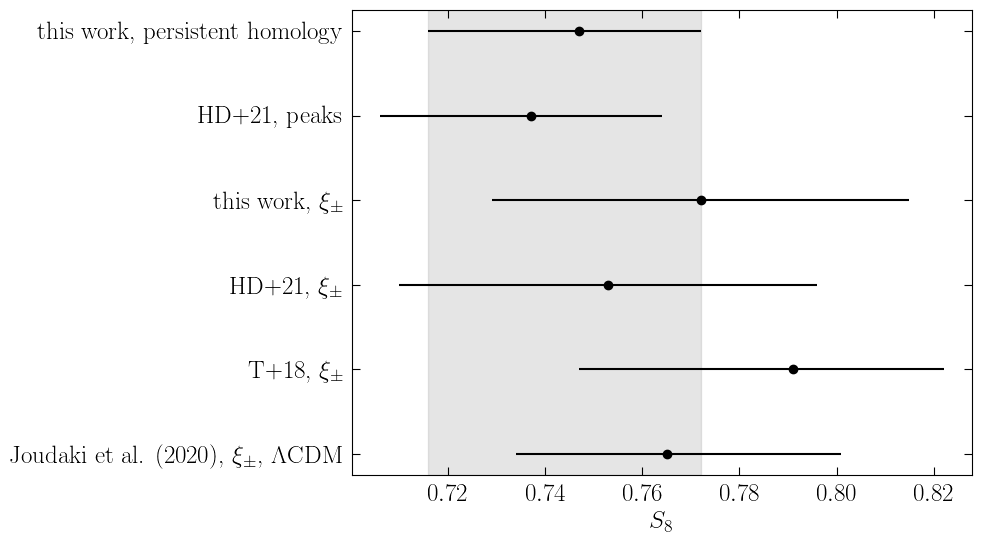}}
    \caption{A comparison of the constraints on the matter clustering parameter $S_8$ from DES-Y1 survey data in a $w$CDM cosmology with fixed neutrino mass.}
    \label{fig:s8_comparisons}
\end{figure}

In this work, we carried out a likelihood analysis on tomographic cosmic shear data using persistent homology, including the marginalisation over systematic effects, %is feasible. Furthermore, using mock data we 
and have shown from simulated data that the posterior constraints can be significantly improved in a joint analysis. While this holds true especially for the intrinsic alignment parameter $A_\mathrm{IA}$ and the equation-of-state of dark energy $w_0$,  the constraints on the matter clustering parameter $S_8$ also improve substantially.

For our analysis we had to make a number of choices, including which persistence statistic to use, which smoothing scale to apply to the heatmaps, and which data compression method to utilise. We have noticed that the posterior constraints achieved by a likelihood analysis do not strongly depend on any of these choices, as can be seen for example in Fig.~\ref{fig:data_compression_comparison}. While further fine-tuning could probably slightly improve the constraining power of our analysis, we believe that this overall stability with respect to different  analysis choices %means that we are not achieving our constraints by utilising numerical (or noise-) artefacts, but rather due to the 
provides strong evidence that we have reached the true sensitivity of persistence statistics to cosmology.

When applying our analysis pipeline to real data, we find that high values of the matter density parameter $\Omm$ are preferred, as observed in HD+21, where a fully independent pipeline and a different summary statistic were utilised. The remarkable similarity between these results suggests that peak count statistics and persistent homology quantify similar aspects of the large scale structure distribution. The fact that both methods favour larger values of $\Omm$ may point to a statistical fluctuation in the DES-Y1 data, or an unknown effect modifying the topological structure of the matter distribution. We investigate this in App.~\ref{sec:app_tension} and show that the chance of such a tension arising due to a statistical fluctuation in the data is about $0.5\%$. We also note that our underlying training data, i.e. the shear catalogues of the {\it Cosmology Training Set}s, are the same as the ones used in HD+21, so this bias might also point towards a statistical fluctuation in this training set. A larger simulation suite would  be able to shine light on this, even though this seems unlikely given the fact that the tension does not exist when validating on simulated data (See Fig. \ref{fig:mcmc_joint}).

We performed several consistency checks to investigate whether the tension is artificially created by our analysis setup: We tried a parameter inference with two-point correlation functions by measuring them in the {\it Cosmology Training Set} and emulating them via the same pipeline that we used for persistent homology. Furthermore we tried removing some nodes from the {\it Cosmology Training Set} and applying different methods of data compression. The results were stable under all these tests, suggesting that the simulation-based inference is not the driver of the tension we measure for $\Omm$.

Another possible explanation is that the DES-Y1 data include an effect that we have not accounted for. This might be an unknown systematic, or a sign of new physics. For example, 2PCF are not sensitive to primordial non-Gaussianities, whereas persistent homology is \citep{Biagetti:2022}.

Overall our constraints on $S_8$ are consistent with previous works (see Fig.~\ref{fig:s8_comparisons}), the largest discrepancy is between our analysis and the one from T+18, which is mainly driven by a different method of estimating the source redshift distribution.

When comparing our results to similar works with peak count statistics, the constraining power of persistent homology appears to be slightly better. In addition, there are a few key differences between our work and HD+21. Firstly, while HD+21 apply a boost-factor to account for baryons, we marginalise over a continuous baryonic effects (and intrinsic alignments) with a wide prior, which is inflating our constraints. Secondly, we account for the emulator uncertainty as described in Sect.~\ref{sec:cosmological_parameter_estimation}; this is not done in HD+21. Comparing with H+21, we see that this emulator uncertainty also inflates cosmological parameter constraints, indicating that the contours reported in HD+21 may be slightly too small. This effect is amplified by the fact that we were only able to train our emulator on 9 lines-of-sight per cosmology, compared to the 50 lines-of-sight in H+21.
Lastly, and most importantly, we have shown in H+21 that persistent homology excels in a high signal-to-noise range, which is not accessible in a tomographic analysis of current-generation surveys. We thus expect this method to out-perform several other higher-order statistics in next-generation surveys.

\begin{acknowledgements}
SH acknowledges support from the German Research Foundation (DFG SCHN 342/13), the International Max-Planck Research School (IMPRS) and the German Academic Scholarship Foundation.
SU acknowledges support from the Max Planck Society and the Alexander von Humboldt Foundation in the framework of the Max Planck-Humboldt Research Award endowed by the Federal Ministry of Education and Research.
Joachim Harnois-D\'eraps acknowledges support from an STFC Ernest Rutherford Fellowship (project reference ST/S004858/1). Computations for the $N$-body simulations were enabled by Compute Ontario (www.computeontario.ca), Westgrid (www.westgrid.ca) and Compute Canada (www.computecanada.ca). The SLICS numerical simulations can be found at http://slics.roe.ac.uk/, while the {\it cosmo}-SLICS can be made available upon request.
TC is supported by the INFN INDARK PD51 grant and by the fare Miur grant `ClustersXEuclid' R165SBKTMA. KD acknowledges support by the COMPLEX project from the European Research Council (ERC) under the European Union’s Horizon 2020 research and innovation program grant agreement ERC-2019-AdG 882679 the Deutsche Forschungsgemeinschaft (DFG, German Research Foundation) under Germany’s Excellence Strategy - EXC-2094 - 390783311 and by the DFG project nr. 490702358.
The Magneticum Simulations were carried out at the Leibniz Supercomputer Center (LRZ) under the project pr86re and pr83li.

\\
\\
\emph{Author contributions}: All authors contributed to the development and writing of this paper. After the lead author, they are separated into two groups, both listed alphabetically. In the first group, BB developed the necessary mathematical background, PB oversaw the measurement and validation of two-point statistics, JHD provided the suites of numerical simulation tailored to the measurement and SU developed the integration into {\sc cosmoSIS}. In the second group, TC, KD and NM were responsible for running and post-processing the Magneticum simulations, extracting the mass maps and computing the lensing statistics.
\end{acknowledgements}

\bibliographystyle{aa}
\bibliography{cite}

\begin{thebibliography}{102}
\expandafter\ifx\csname natexlab\endcsname\relax\def\natexlab#1{#1}\fi

\bibitem[{{Abbott} {et~al.}(2018){Abbott}, {Abdalla}, {Allam}, {Amara},
  {Annis}, {Asorey}, {Avila}, {Ballester}, {Banerji}, {Barkhouse}, {Baruah},
  {Baumer}, {Bechtol}, {Becker}, {Benoit-L{\'e}vy}, {Bernstein}, {Bertin},
  {Blazek}, {Bocquet}, {Brooks}, {Brout}, {Buckley-Geer}, {Burke}, {Busti},
  {Campisano}, {Cardiel-Sas}, {Carnero Rosell}, {Carrasco Kind}, {Carretero},
  {Castander}, {Cawthon}, {Chang}, {Chen}, {Conselice}, {Costa}, {Crocce},
  {Cunha}, {D'Andrea}, {da Costa}, {Das}, {Daues}, {Davis}, {Davis}, {De
  Vicente}, {DePoy}, {DeRose}, {Desai}, {Diehl}, {Dietrich}, {Dodelson},
  {Doel}, {Drlica-Wagner}, {Eifler}, {Elliott}, {Evrard}, {Farahi}, {Fausti
  Neto}, {Fernandez}, {Finley}, {Flaugher}, {Foley}, {Fosalba}, {Friedel},
  {Frieman}, {Garc{\'\i}a-Bellido}, {Gaztanaga}, {Gerdes}, {Giannantonio},
  {Gill}, {Glazebrook}, {Goldstein}, {Gower}, {Gruen}, {Gruendl}, {Gschwend},
  {Gupta}, {Gutierrez}, {Hamilton}, {Hartley}, {Hinton}, {Hislop}, {Hollowood},
  {Honscheid}, {Hoyle}, {Huterer}, {Jain}, {James}, {Jeltema}, {Johnson},
  {Johnson}, {Kacprzak}, {Kent}, {Khullar}, {Klein}, {Kovacs}, {Koziol},
  {Krause}, {Kremin}, {Kron}, {Kuehn}, {Kuhlmann}, {Kuropatkin}, {Lahav},
  {Lasker}, {Li}, {Li}, {Liddle}, {Lima}, {Lin}, {L{\'o}pez-Reyes}, {MacCrann},
  {Maia}, {Maloney}, {Manera}, {March}, {Marriner}, {Marshall}, {Martini},
  {McClintock}, {McKay}, {McMahon}, {Melchior}, {Menanteau}, {Miller},
  {Miquel}, {Mohr}, {Morganson}, {Mould}, {Neilsen}, {Nichol}, {Nogueira},
  {Nord}, {Nugent}, {Nunes}, {Ogand o}, {Old}, {Pace}, {Palmese},
  {Paz-Chinch{\'o}n}, {Peiris}, {Percival}, {Petravick}, {Plazas}, {Poh},
  {Pond}, {Porredon}, {Pujol}, {Refregier}, {Reil}, {Ricker}, {Rollins},
  {Romer}, {Roodman}, {Rooney}, {Ross}, {Rykoff}, {Sako}, {Sanchez}, {Sanchez},
  {Santiago}, {Saro}, {Scarpine}, {Scolnic}, {Serrano}, {Sevilla-Noarbe},
  {Sheldon}, {Shipp}, {Silveira}, {Smith}, {Smith}, {Smith}, {Soares-Santos},
  {Sobreira}, {Song}, {Stebbins}, {Suchyta}, {Sullivan}, {Swanson}, {Tarle},
  {Thaler}, {Thomas}, {Thomas}, {Troxel}, {Tucker}, {Vikram}, {Vivas},
  {Walker}, {Wechsler}, {Weller}, {Wester}, {Wolf}, {Wu}, {Yanny}, {Zenteno},
  {Zhang}, {Zuntz}, {DES Collaboration}, {Juneau}, {Fitzpatrick}, {Nikutta},
  {Nidever}, {Olsen}, {Scott}, \& {NOAO Data Lab}}]{DESY1_data}
{Abbott}, T.~M.~C., {Abdalla}, F.~B., {Allam}, S., {et~al.} 2018, \apjs, 239,
  18

\bibitem[{{Aihara} {et~al.}(2018){Aihara}, {Arimoto}, {Armstrong}, {Arnouts},
  {Bahcall}, {Bickerton}, {Bosch}, {Bundy}, {Capak}, {Chan}, {Chiba}, {Coupon},
  {Egami}, {Enoki}, {Finet}, {Fujimori}, {Fujimoto}, {Furusawa}, {Furusawa},
  {Goto}, {Goulding}, {Greco}, {Greene}, {Gunn}, {Hamana}, {Harikane},
  {Hashimoto}, {Hattori}, {Hayashi}, {Hayashi}, {He{\l}miniak}, {Higuchi},
  {Hikage}, {Ho}, {Hsieh}, {Huang}, {Huang}, {Ikeda}, {Imanishi}, {Inoue},
  {Iwasawa}, {Iwata}, {Jaelani}, {Jian}, {Kamata}, {Karoji}, {Kashikawa},
  {Katayama}, {Kawanomoto}, {Kayo}, {Koda}, {Koike}, {Kojima}, {Komiyama},
  {Konno}, {Koshida}, {Koyama}, {Kusakabe}, {Leauthaud}, {Lee}, {Lin}, {Lin},
  {Lupton}, {Mandelbaum}, {Matsuoka}, {Medezinski}, {Mineo}, {Miyama},
  {Miyatake}, {Miyazaki}, {Momose}, {More}, {More}, {Moritani}, {Moriya},
  {Morokuma}, {Mukae}, {Murata}, {Murayama}, {Nagao}, {Nakata}, {Niida},
  {Niikura}, {Nishizawa}, {Obuchi}, {Oguri}, {Oishi}, {Okabe}, {Okamoto},
  {Okura}, {Ono}, {Onodera}, {Onoue}, {Osato}, {Ouchi}, {Price}, {Pyo}, {Sako},
  {Sawicki}, {Shibuya}, {Shimasaku}, {Shimono}, {Shirasaki}, {Silverman},
  {Simet}, {Speagle}, {Spergel}, {Strauss}, {Sugahara}, {Sugiyama}, {Suto},
  {Suyu}, {Suzuki}, {Tait}, {Takada}, {Takata}, {Tamura}, {Tanaka}, {Tanaka},
  {Tanaka}, {Tanaka}, {Terai}, {Terashima}, {Toba}, {Tominaga}, {Toshikawa},
  {Turner}, {Uchida}, {Uchiyama}, {Umetsu}, {Uraguchi}, {Urata}, {Usuda},
  {Utsumi}, {Wang}, {Wang}, {Wong}, {Yabe}, {Yamada}, {Yamanoi}, {Yasuda},
  {Yeh}, {Yonehara}, \& {Yuma}}]{aihara2018}
{Aihara}, H., {Arimoto}, N., {Armstrong}, R., {et~al.} 2018, \pasj, 70, S4

\bibitem[{{Asgari} {et~al.}(2021){Asgari}, {Lin}, {Joachimi}, {Giblin},
  {Heymans}, {Hildebrandt}, {Kannawadi}, {St{\"o}lzner}, {Tr{\"o}ster}, {van
  den Busch}, {Wright}, {Bilicki}, {Blake}, {de Jong}, {Dvornik}, {Erben},
  {Getman}, {Hoekstra}, {K{\"o}hlinger}, {Kuijken}, {Miller}, {Radovich},
  {Schneider}, {Shan}, \& {Valentijn}}]{asgari2021}
{Asgari}, M., {Lin}, C.-A., {Joachimi}, B., {et~al.} 2021, \aap, 645, A104

\bibitem[{{Asgari} \& {Schneider}(2015)}]{Asgari:2015}
{Asgari}, M. \& {Schneider}, P. 2015, \aap, 578, A50

\bibitem[{{Asgari} {et~al.}(2020){Asgari}, {Tr{\"o}ster}, {Heymans},
  {Hildebrandt}, {van den Busch}, {Wright}, {Choi}, {Erben}, {Joachimi},
  {Joudaki}, {Kannawadi}, {Kuijken}, {Lin}, {Schneider}, \&
  {Zuntz}}]{Asgari:2020}
{Asgari}, M., {Tr{\"o}ster}, T., {Heymans}, C., {et~al.} 2020, \aap, 634, A127

\bibitem[{{Bartelmann} \& {Schneider}(2001)}]{bartelmann:2001}
{Bartelmann}, M. \& {Schneider}, P. 2001, \physrep, 340, 291

\bibitem[{{Ben{\'{\i}}tez}(2000)}]{BPZ}
{Ben{\'{\i}}tez}, N. 2000, \apj, 536, 571

\bibitem[{{Berg{\'e}} {et~al.}(2010){Berg{\'e}}, {Amara}, \&
  {R{\'e}fr{\'e}gier}}]{Berge2010}
{Berg{\'e}}, J., {Amara}, A., \& {R{\'e}fr{\'e}gier}, A. 2010, \apj, 712, 992

\bibitem[{{Biagetti} {et~al.}(2022){Biagetti}, {Calles}, {Castiblanco}, {Cole},
  \& {Nore{\~n}a}}]{Biagetti:2022}
{Biagetti}, M., {Calles}, J., {Castiblanco}, L., {Cole}, A., \& {Nore{\~n}a},
  J. 2022, arXiv e-prints, arXiv:2203.08262

\bibitem[{{Biagetti} {et~al.}(2020){Biagetti}, {Cole}, \&
  {Shiu}}]{biagetti2020}
{Biagetti}, M., {Cole}, A., \& {Shiu}, G. 2020, arXiv e-prints,
  arXiv:2009.04819

\bibitem[{{Biffi} {et~al.}(2013){Biffi}, {Dolag}, \&
  {B{\"o}hringer}}]{Biffi:2013}
{Biffi}, V., {Dolag}, K., \& {B{\"o}hringer}, H. 2013, \mnras, 428, 1395

\bibitem[{{Blazek} {et~al.}(2019){Blazek}, {MacCrann}, {Troxel}, \&
  {Fang}}]{Blazek2019}
{Blazek}, J.~A., {MacCrann}, N., {Troxel}, M.~A., \& {Fang}, X. 2019, \prd,
  100, 103506

\bibitem[{{Bocquet} {et~al.}(2016){Bocquet}, {Saro}, {Dolag}, \&
  {Mohr}}]{Bocquet:2016}
{Bocquet}, S., {Saro}, A., {Dolag}, K., \& {Mohr}, J.~J. 2016, \mnras, 456,
  2361

\bibitem[{{Bridle} \& {King}(2007)}]{BridleKing}
{Bridle}, S. \& {King}, L. 2007, New Journal of Physics, 9, 444

\bibitem[{Bubenik(2015)}]{Bub:Statisticaltopologicaldata}
Bubenik, P. 2015, Journal of Machine Learning Research (JMLR), 16, 77

\bibitem[{{Burger} {et~al.}(2021){Burger}, {Friedrich}, {Harnois-D{\'e}raps},
  \& {Schneider}}]{Burger2021}
{Burger}, P., {Friedrich}, O., {Harnois-D{\'e}raps}, J., \& {Schneider}, P.
  2021, arXiv e-prints, arXiv:2106.13214

\bibitem[{{Castro} {et~al.}(2021){Castro}, {Borgani}, {Dolag}, {Marra},
  {Quartin}, {Saro}, \& {Sefusatti}}]{MagneticumBox2b}
{Castro}, T., {Borgani}, S., {Dolag}, K., {et~al.} 2021, \mnras, 500, 2316

\bibitem[{{Castro} {et~al.}(2018{\natexlab{a}}){Castro}, {Quartin}, {Giocoli},
  {Borgani}, \& {Dolag}}]{Castro:2018}
{Castro}, T., {Quartin}, M., {Giocoli}, C., {Borgani}, S., \& {Dolag}, K.
  2018{\natexlab{a}}, \mnras, 478, 1305

\bibitem[{{Castro} {et~al.}(2018{\natexlab{b}}){Castro}, {Quartin}, {Giocoli},
  {Borgani}, \& {Dolag}}]{LensingPDF_baryons}
{Castro}, T., {Quartin}, M., {Giocoli}, C., {Borgani}, S., \& {Dolag}, K.
  2018{\natexlab{b}}, \mnras, 478, 1305

\bibitem[{Chazal \& Michel(2021)}]{CM:introductionTopologicalData}
Chazal, F. \& Michel, B. 2021, Frontiers in Artificial Intelligence, 4, 108

\bibitem[{Chittajallu {et~al.}(2018)Chittajallu, Siekierski, Lee, Gerber,
  Beezley, Manthey, Gutman, \& Cooper}]{CSL+:Vectorizedpersistenthomology}
Chittajallu, D.~R., Siekierski, N., Lee, S., {et~al.} 2018, in 2018 IEEE 15th
  International Symposium on Biomedical Imaging (ISBI 2018), IEEE, 232--235

\bibitem[{{Coulton} {et~al.}(2020){Coulton}, {Liu}, {McCarthy}, \&
  {Osato}}]{MinimaPeaks}
{Coulton}, W.~R., {Liu}, J., {McCarthy}, I.~G., \& {Osato}, K. 2020, \mnras,
  495, 2531

\bibitem[{{de Jong} {et~al.}(2013){de Jong}, {Verdoes Kleijn}, {Kuijken}, \&
  {Valentijn}}]{dejong2013}
{de Jong}, J. T.~A., {Verdoes Kleijn}, G.~A., {Kuijken}, K.~H., \& {Valentijn},
  E.~A. 2013, Experimental Astronomy, 35, 25

\bibitem[{{DES Collaboration} {et~al.}(2021){DES Collaboration}, {Abbott},
  {Aguena}, {Alarcon}, {Allam}, {Alves}, {Amon}, {Andrade-Oliveira}, {Annis},
  {Avila}, {Bacon}, {Baxter}, {Bechtol}, {Becker}, {Bernstein}, {Bhargava},
  {Birrer}, {Blazek}, {Brandao-Souza}, {Bridle}, {Brooks}, {Buckley-Geer},
  {Burke}, {Camacho}, {Campos}, {Carnero Rosell}, {Carrasco Kind}, {Carretero},
  {Castander}, {Cawthon}, {Chang}, {Chen}, {Chen}, {Choi}, {Conselice},
  {Cordero}, {Costanzi}, {Crocce}, {da Costa}, {da Silva Pereira}, {Davis},
  {Davis}, {De Vicente}, {DeRose}, {Desai}, {Di Valentino}, {Diehl},
  {Dietrich}, {Dodelson}, {Doel}, {Doux}, {Drlica-Wagner}, {Eckert}, {Eifler},
  {Elsner}, {Elvin-Poole}, {Everett}, {Evrard}, {Fang}, {Farahi}, {Fernandez},
  {Ferrero}, {Fert{\'e}}, {Fosalba}, {Friedrich}, {Frieman},
  {Garc{\'\i}a-Bellido}, {Gatti}, {Gaztanaga}, {Gerdes}, {Giannantonio},
  {Giannini}, {Gruen}, {Gruendl}, {Gschwend}, {Gutierrez}, {Harrison},
  {Hartley}, {Herner}, {Hinton}, {Hollowood}, {Honscheid}, {Hoyle}, {Huff},
  {Huterer}, {Jain}, {James}, {Jarvis}, {Jeffrey}, {Jeltema}, {Kovacs},
  {Krause}, {Kron}, {Kuehn}, {Kuropatkin}, {Lahav}, {Leget}, {Lemos}, {Liddle},
  {Lidman}, {Lima}, {Lin}, {MacCrann}, {Maia}, {Marshall}, {Martini},
  {McCullough}, {Melchior}, {Mena-Fern{\'a}ndez}, {Menanteau}, {Miquel},
  {Mohr}, {Morgan}, {Muir}, {Myles}, {Nadathur}, {Navarro-Alsina}, {Nichol},
  {Ogando}, {Omori}, {Palmese}, {Pandey}, {Park}, {Paz-Chinch{\'o}n},
  {Petravick}, {Pieres}, {Plazas Malag{\'o}n}, {Porredon}, {Prat}, {Raveri},
  {Rodriguez-Monroy}, {Rollins}, {Romer}, {Roodman}, {Rosenfeld}, {Ross},
  {Rykoff}, {Samuroff}, {S{\'a}nchez}, {Sanchez}, {Sanchez}, {Sanchez Cid},
  {Scarpine}, {Schubnell}, {Scolnic}, {Secco}, {Serrano}, {Sevilla-Noarbe},
  {Sheldon}, {Shin}, {Smith}, {Soares-Santos}, {Suchyta}, {Swanson}, {Tabbutt},
  {Tarle}, {Thomas}, {To}, {Troja}, {Troxel}, {Tucker}, {Tutusaus}, {Varga},
  {Walker}, {Weaverdyck}, {Weller}, {Yanny}, {Yin}, {Zhang}, \&
  {Zuntz}}]{DES2021}
{DES Collaboration}, {Abbott}, T.~M.~C., {Aguena}, M., {et~al.} 2021, arXiv
  e-prints, arXiv:2105.13549

\bibitem[{Dlotko(2020)}]{gudhi:CubicalComplex}
Dlotko, P. 2020, in {GUDHI} User and Reference Manual, {3.1.1} edn. ({GUDHI
  Editorial Board})

\bibitem[{{Dolag}(2015)}]{Dolag:2015}
{Dolag}, K. 2015, in IAU General Assembly, Vol.~29, 2250156

\bibitem[{{Feldbrugge} {et~al.}(2019){Feldbrugge}, {van Engelen}, {van de
  Weygaert}, {Pranav}, \& {Vegter}}]{feldbrugge2019}
{Feldbrugge}, J., {van Engelen}, M., {van de Weygaert}, R., {Pranav}, P., \&
  {Vegter}, G. 2019, \jcap, 2019, 052

\bibitem[{{Ferreira} {et~al.}(2021){Ferreira}, {Zhang}, {Chen}, {Dodelson}, \&
  {LSST Dark Energy Science Collaboration}}]{Ferreira:2021}
{Ferreira}, T., {Zhang}, T., {Chen}, N., {Dodelson}, S., \& {LSST Dark Energy
  Science Collaboration}. 2021, \prd, 103, 103535

\bibitem[{{Flaugher}(2005)}]{flaugher2005}
{Flaugher}, B. 2005, International Journal of Modern Physics A, 20, 3121

\bibitem[{{Flaugher} {et~al.}(2015){Flaugher}, {Diehl}, {Honscheid}, {Abbott},
  {Alvarez}, {Angstadt}, {Annis}, {Antonik}, {Ballester}, {Beaufore},
  {Bernstein}, {Bernstein}, {Bigelow}, {Bonati}, {Boprie}, {Brooks},
  {Buckley-Geer}, {Campa}, {Cardiel-Sas}, {Castand er}, {Castilla}, {Cease},
  {Cela-Ruiz}, {Chappa}, {Chi}, {Cooper}, {da Costa}, {Dede}, {Derylo},
  {DePoy}, {de Vicente}, {Doel}, {Drlica-Wagner}, {Eiting}, {Elliott}, {Emes},
  {Estrada}, {Fausti Neto}, {Finley}, {Flores}, {Frieman}, {Gerdes},
  {Gladders}, {Gregory}, {Gutierrez}, {Hao}, {Holland}, {Holm}, {Huffman},
  {Jackson}, {James}, {Jonas}, {Karcher}, {Karliner}, {Kent}, {Kessler},
  {Kozlovsky}, {Kron}, {Kubik}, {Kuehn}, {Kuhlmann}, {Kuk}, {Lahav}, {Lathrop},
  {Lee}, {Levi}, {Lewis}, {Li}, {Mand richenko}, {Marshall}, {Martinez},
  {Merritt}, {Miquel}, {Mu{\~n}oz}, {Neilsen}, {Nichol}, {Nord}, {Ogando},
  {Olsen}, {Palaio}, {Patton}, {Peoples}, {Plazas}, {Rauch}, {Reil}, {Rheault},
  {Roe}, {Rogers}, {Roodman}, {Sanchez}, {Scarpine}, {Schindler}, {Schmidt},
  {Schmitt}, {Schubnell}, {Schultz}, {Schurter}, {Scott}, {Serrano}, {Shaw},
  {Smith}, {Soares-Santos}, {Stefanik}, {Stuermer}, {Suchyta}, {Sypniewski},
  {Tarle}, {Thaler}, {Tighe}, {Tran}, {Tucker}, {Walker}, {Wang}, {Watson},
  {Weaverdyck}, {Wester}, {Woods}, {Yanny}, \& {DES Collaboration}}]{DES_CAM}
{Flaugher}, B., {Diehl}, H.~T., {Honscheid}, K., {et~al.} 2015, \aj, 150, 150

\bibitem[{{Gruen} \& {Brimioulle}(2017)}]{GruenBrimioulle}
{Gruen}, D. \& {Brimioulle}, F. 2017, \mnras, 468, 769

\bibitem[{{Halder} {et~al.}(2021){Halder}, {Friedrich}, {Seitz}, \&
  {Varga}}]{Halder2021}
{Halder}, A., {Friedrich}, O., {Seitz}, S., \& {Varga}, T.~N. 2021, \mnras,
  506, 2780

\bibitem[{{Hamana} {et~al.}(2020){Hamana}, {Shirasaki}, {Miyazaki}, {Hikage},
  {Oguri}, {More}, {Armstrong}, {Leauthaud}, {Mandelbaum}, {Miyatake},
  {Nishizawa}, {Simet}, {Takada}, {Aihara}, {Bosch}, {Komiyama}, {Lupton},
  {Murayama}, {Strauss}, \& {Tanaka}}]{Hamana:2020}
{Hamana}, T., {Shirasaki}, M., {Miyazaki}, S., {et~al.} 2020, \pasj, 72, 16

\bibitem[{{Handley} {et~al.}(2015){Handley}, {Hobson}, \&
  {Lasenby}}]{Handley:2015}
{Handley}, W.~J., {Hobson}, M.~P., \& {Lasenby}, A.~N. 2015, \mnras, 453, 4384

\bibitem[{{Harnois-D{\'e}raps} {et~al.}(2019){Harnois-D{\'e}raps}, {Giblin}, \&
  {Joachimi}}]{harnois-deraps2019}
{Harnois-D{\'e}raps}, J., {Giblin}, B., \& {Joachimi}, B. 2019, \aap, 631, A160

\bibitem[{{Harnois-D{\'e}raps} {et~al.}(2021){Harnois-D{\'e}raps}, {Martinet},
  {Castro}, {Dolag}, {Giblin}, {Heymans}, {Hildebrandt}, \&
  {Xia}}]{harnois-deraps:2021}
{Harnois-D{\'e}raps}, J., {Martinet}, N., {Castro}, T., {et~al.} 2021, \mnras,
  506, 1623

\bibitem[{{Harnois-D{\'e}raps} {et~al.}(2022){Harnois-D{\'e}raps}, {Martinet},
  \& {Reischke}}]{JHD21b}
{Harnois-D{\'e}raps}, J., {Martinet}, N., \& {Reischke}, R. 2022, \mnras, 509,
  3868

\bibitem[{{Harnois-D{\'e}raps} {et~al.}(2013){Harnois-D{\'e}raps}, {Pen},
  {Iliev}, {Merz}, {Emberson}, \& {Desjacques}}]{CUBEP3M}
{Harnois-D{\'e}raps}, J., {Pen}, U.-L., {Iliev}, I.~T., {et~al.} 2013, \mnras,
  436, 540

\bibitem[{{Harnois-D{\'e}raps} \& {van Waerbeke}(2015)}]{harnois-deraps2015}
{Harnois-D{\'e}raps}, J. \& {van Waerbeke}, L. 2015, \mnras, 450, 2857

\bibitem[{{Hartlap} {et~al.}(2007){Hartlap}, {Simon}, \&
  {Schneider}}]{Hartlap:2007}
{Hartlap}, J., {Simon}, P., \& {Schneider}, P. 2007, \aap, 464, 399

\bibitem[{{Heavens} {et~al.}(2000){Heavens}, {Jimenez}, \&
  {Lahav}}]{Heavens:2000}
{Heavens}, A.~F., {Jimenez}, R., \& {Lahav}, O. 2000, \mnras, 317, 965

\bibitem[{{Heavens} {et~al.}(2017){Heavens}, {Sellentin}, {de Mijolla}, \&
  {Vianello}}]{Heavens:2017}
{Heavens}, A.~F., {Sellentin}, E., {de Mijolla}, D., \& {Vianello}, A. 2017,
  \mnras, 472, 4244

\bibitem[{{Hetterscheidt} {et~al.}(2005){Hetterscheidt}, {Erben}, {Schneider},
  {Maoli}, {van Waerbeke}, \& {Mellier}}]{Hetterscheidt:2005}
{Hetterscheidt}, M., {Erben}, T., {Schneider}, P., {et~al.} 2005, \aap, 442, 43

\bibitem[{{Heydenreich} {et~al.}(2021){Heydenreich}, {Br{\"u}ck}, \&
  {Harnois-D{\'e}raps}}]{heydenreich2021}
{Heydenreich}, S., {Br{\"u}ck}, B., \& {Harnois-D{\'e}raps}, J. 2021, \aap,
  648, A74

\bibitem[{{Heymans} {et~al.}(2021){Heymans}, {Tr{\"o}ster}, {Asgari}, {Blake},
  {Hildebrandt}, {Joachimi}, {Kuijken}, {Lin}, {S{\'a}nchez}, {van den Busch},
  {Wright}, {Amon}, {Bilicki}, {de Jong}, {Crocce}, {Dvornik}, {Erben},
  {Fortuna}, {Getman}, {Giblin}, {Glazebrook}, {Hoekstra}, {Joudaki},
  {Kannawadi}, {K{\"o}hlinger}, {Lidman}, {Miller}, {Napolitano}, {Parkinson},
  {Schneider}, {Shan}, {Valentijn}, {Verdoes Kleijn}, \& {Wolf}}]{heymans2021}
{Heymans}, C., {Tr{\"o}ster}, T., {Asgari}, M., {et~al.} 2021, \aap, 646, A140

\bibitem[{{Hikage} {et~al.}(2019){Hikage}, {Oguri}, {Hamana}, {More},
  {Mandelbaum}, {Takada}, {K{\"o}hlinger}, {Miyatake}, {Nishizawa}, {Aihara},
  {Armstrong}, {Bosch}, {Coupon}, {Ducout}, {Ho}, {Hsieh}, {Komiyama},
  {Lanusse}, {Leauthaud}, {Lupton}, {Medezinski}, {Mineo}, {Miyama},
  {Miyazaki}, {Murata}, {Murayama}, {Shirasaki}, {Sif{\'o}n}, {Simet},
  {Speagle}, {Spergel}, {Strauss}, {Sugiyama}, {Tanaka}, {Utsumi}, {Wang}, \&
  {Yamada}}]{hikage2019}
{Hikage}, C., {Oguri}, M., {Hamana}, T., {et~al.} 2019, \pasj, 71, 43

\bibitem[{{Hildebrandt} {et~al.}(2018){Hildebrandt}, {K{\"o}hlinger}, {van den
  Busch}, {Joachimi}, {Heymans}, {Kannawadi}, {Wright}, {Asgari}, {Blake},
  {Hoekstra}, {Joudaki}, {Kuijken}, {Miller}, {Morrison}, {Tr{\"o}ster},
  {Amon}, {Archidiacono}, {Brieden}, {Choi}, {de Jong}, {Erben}, {Giblin},
  {Mead}, {Peacock}, {Radovich}, {Schneider}, {Sif{\'o}n}, \& {Tewes}}]{KV450}
{Hildebrandt}, H., {K{\"o}hlinger}, F., {van den Busch}, J.~L., {et~al.} 2018,
  arXiv e-prints, arXiv:1812.06076

\bibitem[{{Hildebrandt} {et~al.}(2017){Hildebrandt}, {Viola}, {Heymans},
  {Joudaki}, {Kuijken}, {Blake}, {Erben}, {Joachimi}, {Klaes}, {Miller},
  {Morrison}, {Nakajima}, {Verdoes Kleijn}, {Amon}, {Choi}, {Covone}, {de
  Jong}, {Dvornik}, {Fenech Conti}, {Grado}, {Harnois-D{\'e}raps}, {Herbonnet},
  {Hoekstra}, {K{\"o}hlinger}, {McFarland}, {Mead}, {Merten}, {Napolitano},
  {Peacock}, {Radovich}, {Schneider}, {Simon}, {Valentijn}, {van den Busch},
  {van Uitert}, \& {Van Waerbeke}}]{hildebrandt2017}
{Hildebrandt}, H., {Viola}, M., {Heymans}, C., {et~al.} 2017, \mnras, 465, 1454

\bibitem[{{Hirschmann} {et~al.}(2014){Hirschmann}, {Dolag}, {Saro}, {Bachmann},
  {Borgani}, \& {Burkert}}]{2014MNRAS.442.2304H}
{Hirschmann}, M., {Dolag}, K., {Saro}, A., {et~al.} 2014, \mnras, 442, 2304

\bibitem[{{Hoyle} {et~al.}(2018){Hoyle}, {Gruen}, {Bernstein}, {Rau}, {De
  Vicente}, {Hartley}, {Gaztanaga}, {DeRose}, {Troxel}, {Davis}, {Alarcon},
  {MacCrann}, {Prat}, {S{\'a}nchez}, {Sheldon}, {Wechsler}, {Asorey}, {Becker},
  {Bonnett}, {Carnero Rosell}, {Carollo}, {Carrasco Kind}, {Castander},
  {Cawthon}, {Chang}, {Childress}, {Davis}, {Drlica-Wagner}, {Gatti},
  {Glazebrook}, {Gschwend}, {Hinton}, {Hoormann}, {Kim}, {King}, {Kuehn},
  {Lewis}, {Lidman}, {Lin}, {Macaulay}, {Maia}, {Martini}, {Mudd},
  {M{\"o}ller}, {Nichol}, {Ogando}, {Rollins}, {Roodman}, {Ross}, {Rozo},
  {Rykoff}, {Samuroff}, {Sevilla-Noarbe}, {Sharp}, {Sommer}, {Tucker}, {Uddin},
  {Varga}, {Vielzeuf}, {Yuan}, {Zhang}, {Abbott}, {Abdalla}, {Allam}, {Annis},
  {Bechtol}, {Benoit-L{\'e}vy}, {Bertin}, {Brooks}, {Buckley-Geer}, {Burke},
  {Busha}, {Capozzi}, {Carretero}, {Crocce}, {D'Andrea}, {da Costa}, {DePoy},
  {Desai}, {Diehl}, {Doel}, {Eifler}, {Estrada}, {Evrard}, {Fernandez},
  {Flaugher}, {Fosalba}, {Frieman}, {Garc{\'\i}a-Bellido}, {Gerdes},
  {Giannantonio}, {Goldstein}, {Gruendl}, {Gutierrez}, {Honscheid}, {James},
  {Jarvis}, {Jeltema}, {Johnson}, {Johnson}, {Kirk}, {Krause}, {Kuhlmann},
  {Kuropatkin}, {Lahav}, {Li}, {Lima}, {March}, {Marshall}, {Melchior},
  {Menanteau}, {Miquel}, {Nord}, {O'Neill}, {Plazas}, {Romer}, {Sako},
  {Sanchez}, {Santiago}, {Scarpine}, {Schindler}, {Schubnell}, {Smith},
  {Smith}, {Soares-Santos}, {Sobreira}, {Suchyta}, {Swanson}, {Tarle},
  {Thomas}, {Tucker}, {Vikram}, {Walker}, {Weller}, {Wester}, {Wolf}, {Yanny},
  {Zuntz}, \& {DES Collaboration}}]{DESY1_redshifts}
{Hoyle}, B., {Gruen}, D., {Bernstein}, G.~M., {et~al.} 2018, \mnras, 478, 592

\bibitem[{{Ivezic} {et~al.}(2008){Ivezic}, {Axelrod}, {Brandt}, {Burke},
  {Claver}, {Connolly}, {Cook}, {Gee}, {Gilmore}, {Jacoby}, {Jones}, {Kahn},
  {Kantor}, {Krabbendam}, {Lupton}, {Monet}, {Pinto}, {Saha}, {Schalk},
  {Schneider}, {Strauss}, {Stubbs}, {Sweeney}, {Szalay}, {Thaler}, {Tyson}, \&
  {LSST Collaboration}}]{ivezic2008}
{Ivezic}, Z., {Axelrod}, T., {Brandt}, W.~N., {et~al.} 2008, Serbian
  Astronomical Journal, 176, 1

\bibitem[{{Jarvis} {et~al.}(2004){Jarvis}, {Bernstein}, \&
  {Jain}}]{Jarvis:Bernstein:2004}
{Jarvis}, M., {Bernstein}, G., \& {Jain}, B. 2004, \mnras, 352, 338

\bibitem[{{Joachimi} {et~al.}(2015){Joachimi}, {Cacciato}, {Kitching},
  {Leonard}, {Mandelbaum}, {Sch{\"a}fer}, {Sif{\'o}n}, {Hoekstra}, {Kiessling},
  {Kirk}, \& {Rassat}}]{Joachimi_IA_review}
{Joachimi}, B., {Cacciato}, M., {Kitching}, T.~D., {et~al.} 2015, \ssr, 193, 1

\bibitem[{{Joudaki} {et~al.}(2020){Joudaki}, {Hildebrandt}, {Traykova},
  {Chisari}, {Heymans}, {Kannawadi}, {Kuijken}, {Wright}, {Asgari}, {Erben},
  {Hoekstra}, {Joachimi}, {Miller}, {Tr{\"o}ster}, \& {van den
  Busch}}]{joudaki2020}
{Joudaki}, S., {Hildebrandt}, H., {Traykova}, D., {et~al.} 2020, \aap, 638, L1

\bibitem[{{Joudaki} {et~al.}(2017){Joudaki}, {Mead}, {Blake}, {Choi}, {de
  Jong}, {Erben}, {Fenech Conti}, {Herbonnet}, {Heymans}, {Hildebrandt},
  {Hoekstra}, {Joachimi}, {Klaes}, {K{\"o}hlinger}, {Kuijken}, {McFarland},
  {Miller}, {Schneider}, \& {Viola}}]{Joudaki:2017}
{Joudaki}, S., {Mead}, A., {Blake}, C., {et~al.} 2017, \mnras, 471, 1259

\bibitem[{{Kacprzak} {et~al.}(2016){Kacprzak}, {Kirk}, {Friedrich}, {Amara},
  {Refregier}, {Marian}, {Dietrich}, {Suchyta}, {Aleksi{\'c}}, {Bacon},
  {Becker}, {Bonnett}, {Bridle}, {Chang}, {Eifler}, {Hartley}, {Huff},
  {Krause}, {MacCrann}, {Melchior}, {Nicola}, {Samuroff}, {Sheldon}, {Troxel},
  {Weller}, {Zuntz}, {Abbott}, {Abdalla}, {Armstrong}, {Benoit-L{\'e}vy},
  {Bernstein}, {Bernstein}, {Bertin}, {Brooks}, {Burke}, {Carnero Rosell},
  {Carrasco Kind}, {Carretero}, {Castander}, {Crocce}, {D'Andrea}, {da Costa},
  {Desai}, {Diehl}, {Evrard}, {Neto}, {Flaugher}, {Fosalba}, {Frieman},
  {Gerdes}, {Goldstein}, {Gruen}, {Gruendl}, {Gutierrez}, {Honscheid}, {Jain},
  {James}, {Jarvis}, {Kuehn}, {Kuropatkin}, {Lahav}, {Lima}, {March},
  {Marshall}, {Martini}, {Miller}, {Miquel}, {Mohr}, {Nichol}, {Nord},
  {Plazas}, {Romer}, {Roodman}, {Rykoff}, {Sanchez}, {Scarpine}, {Schubnell},
  {Sevilla-Noarbe}, {Smith}, {Soares-Santos}, {Sobreira}, {Swanson}, {Tarle},
  {Thomas}, {Vikram}, {Walker}, {Zhang}, \& {DES
  Collaboration}}]{Kacprzak:2016}
{Kacprzak}, T., {Kirk}, D., {Friedrich}, O., {et~al.} 2016, \mnras, 463, 3653

\bibitem[{{Kimura} \& {Imai}(2017)}]{kimura2017}
{Kimura}, Y. \& {Imai}, K. 2017, Advances in Space Research, 60, 722

\bibitem[{{Kono} {et~al.}(2020){Kono}, {Takeuchi}, {Cooray}, {Nishizawa}, \&
  {Murakami}}]{kono2020}
{Kono}, K.~T., {Takeuchi}, T.~T., {Cooray}, S., {Nishizawa}, A.~J., \&
  {Murakami}, K. 2020, arXiv e-prints, arXiv:2006.02905

\bibitem[{Kovacev-Nikolic {et~al.}(2016)Kovacev-Nikolic, Bubenik, Nikoli\'{c},
  \& Heo}]{KBNH:Usingpersistenthomology}
Kovacev-Nikolic, V., Bubenik, P., Nikoli\'{c}, D., \& Heo, G. 2016, Statistical
  Applications in Genetics and Molecular Biology, 15, 19

\bibitem[{{Laureijs} {et~al.}(2011){Laureijs}, {Amiaux}, {Arduini},
  {Augu{\`e}res}, {Brinchmann}, {Cole}, {Cropper}, {Dabin}, {Duvet}, {Ealet},
  \& et~al.}]{laureijs2011}
{Laureijs}, R., {Amiaux}, J., {Arduini}, S., {et~al.} 2011, arXiv e-prints,
  arXiv:1110.3193

\bibitem[{{Lemos} {et~al.}(2022){Lemos}, {Weaverdyck}, {Rollins}, {Muir},
  {Fert{\'e}}, {Liddle}, {Campos}, {Huterer}, {Raveri}, {Zuntz}, {Di
  Valentino}, {Fang}, {Hartley}, {Aguena}, {Allam}, {Annis}, {Bertin},
  {Bocquet}, {Brooks}, {Burke}, {Carnero Rosell}, {Carrasco Kind}, {Carretero},
  {Castander}, {Choi}, {Costanzi}, {Crocce}, {da Costa}, {Pereira}, {Dietrich},
  {Everett}, {Ferrero}, {Frieman}, {Garc{\'\i}a-Bellido}, {Gatti}, {Gaztanaga},
  {Gerdes}, {Gruen}, {Gruendl}, {Gschwend}, {Gutierrez}, {Hinton}, {Hollowood},
  {Honscheid}, {James}, {Kuehn}, {Kuropatkin}, {Lima}, {March}, {Melchior},
  {Menanteau}, {Miquel}, {Morgan}, {Palmese}, {Paz-Chinch{\'o}n}, {Pieres},
  {Plazas Malag{\'o}n}, {Porredon}, {Sanchez}, {Scarpine}, {Schubnell},
  {Serrano}, {Sevilla-Noarbe}, {Smith}, {Suchyta}, {Swanson}, {Tarle},
  {Thomas}, {To}, {Varga}, \& {Weller}}]{Lemos:2022}
{Lemos}, P., {Weaverdyck}, N., {Rollins}, R.~P., {et~al.} 2022, arXiv e-prints,
  arXiv:2202.08233

\bibitem[{{Lima} {et~al.}(2008){Lima}, {Cunha}, {Oyaizu}, {Frieman}, {Lin}, \&
  {Sheldon}}]{Lima}
{Lima}, M., {Cunha}, C.~E., {Oyaizu}, H., {et~al.} 2008, \mnras, 390, 118

\bibitem[{{Mandelbaum}(2018)}]{Mandelbaum18}
{Mandelbaum}, R. 2018, \araa, 56, 393

\bibitem[{{Martinet} {et~al.}(2021{\natexlab{a}}){Martinet}, {Castro},
  {Harnois-D{\'e}raps}, {Jullo}, {Giocoli}, \& {Dolag}}]{martinet21b}
{Martinet}, N., {Castro}, T., {Harnois-D{\'e}raps}, J., {et~al.}
  2021{\natexlab{a}}, \aap, 648, A115

\bibitem[{{Martinet} {et~al.}(2021{\natexlab{b}}){Martinet},
  {Harnois-D{\'e}raps}, {Jullo}, \& {Schneider}}]{martinet2021}
{Martinet}, N., {Harnois-D{\'e}raps}, J., {Jullo}, E., \& {Schneider}, P.
  2021{\natexlab{b}}, \aap, 646, A62

\bibitem[{{Martinet} {et~al.}(2018){Martinet}, {Schneider}, {Hildebrandt},
  {Shan}, {Asgari}, {Dietrich}, {Harnois-D{\'e}raps}, {Erben}, {Grado},
  {Heymans}, {Hoekstra}, {Klaes}, {Kuijken}, {Merten}, \&
  {Nakajima}}]{Martinet:2018}
{Martinet}, N., {Schneider}, P., {Hildebrandt}, H., {et~al.} 2018, \mnras, 474,
  712

\bibitem[{{McCarthy} {et~al.}(2017){McCarthy}, {Schaye}, {Bird}, \& {Le
  Brun}}]{BAHAMAS}
{McCarthy}, I.~G., {Schaye}, J., {Bird}, S., \& {Le Brun}, A.~M.~C. 2017,
  \mnras, 465, 2936

\bibitem[{{Navarro} {et~al.}(1997){Navarro}, {Frenk}, \&
  {White}}]{Navarro:1997}
{Navarro}, J.~F., {Frenk}, C.~S., \& {White}, S. D.~M. 1997, \apj, 490, 493

\bibitem[{Otter {et~al.}(2017)Otter, Porter, Tillmann, Grindrod, \&
  Harrington}]{OPT+:roadmapcomputationpersistent}
Otter, N., Porter, M.~A., Tillmann, U., Grindrod, P., \& Harrington, H.~A. 2017
  [\eprint{http://arxiv.org/abs/1506.08903v7}]

\bibitem[{Oudot(2015)}]{Oud:Persistencetheory:quiver}
Oudot, S.~Y. 2015, Mathematical Surveys and Monographs, Vol. 209, Persistence
  theory: from quiver representations to data analysis (American Mathematical
  Society, Providence, RI), viii+218

\bibitem[{{Parroni} {et~al.}(2020){Parroni}, {Cardone}, {Maoli}, \&
  {Scaramella}}]{parroni2020}
{Parroni}, C., {Cardone}, V.~F., {Maoli}, R., \& {Scaramella}, R. 2020, \aap,
  633, A71

\bibitem[{{Petri} {et~al.}(2015){Petri}, {Liu}, {Haiman}, {May}, {Hui}, \&
  {Kratochvil}}]{petri2015}
{Petri}, A., {Liu}, J., {Haiman}, Z., {et~al.} 2015, \prd, 91, 103511

\bibitem[{{Planck Collaboration} {et~al.}(2020){Planck Collaboration},
  {Aghanim}, {Akrami}, {Ashdown}, {Aumont}, {Baccigalupi}, {Ballardini},
  {Banday}, {Barreiro}, {Bartolo}, {Basak}, {Battye}, {Benabed}, {Bernard},
  {Bersanelli}, {Bielewicz}, {Bock}, {Bond}, {Borrill}, {Bouchet}, {Boulanger},
  {Bucher}, {Burigana}, {Butler}, {Calabrese}, {Cardoso}, {Carron},
  {Challinor}, {Chiang}, {Chluba}, {Colombo}, {Combet}, {Contreras}, {Crill},
  {Cuttaia}, {de Bernardis}, {de Zotti}, {Delabrouille}, {Delouis}, {Di
  Valentino}, {Diego}, {Dor{\'e}}, {Douspis}, {Ducout}, {Dupac}, {Dusini},
  {Efstathiou}, {Elsner}, {En{\ss}lin}, {Eriksen}, {Fantaye}, {Farhang},
  {Fergusson}, {Fernandez-Cobos}, {Finelli}, {Forastieri}, {Frailis},
  {Fraisse}, {Franceschi}, {Frolov}, {Galeotta}, {Galli}, {Ganga},
  {G{\'e}nova-Santos}, {Gerbino}, {Ghosh}, {Gonz{\'a}lez-Nuevo}, {G{\'o}rski},
  {Gratton}, {Gruppuso}, {Gudmundsson}, {Hamann}, {Handley}, {Hansen},
  {Herranz}, {Hildebrandt}, {Hivon}, {Huang}, {Jaffe}, {Jones}, {Karakci},
  {Keih{\"a}nen}, {Keskitalo}, {Kiiveri}, {Kim}, {Kisner}, {Knox},
  {Krachmalnicoff}, {Kunz}, {Kurki-Suonio}, {Lagache}, {Lamarre}, {Lasenby},
  {Lattanzi}, {Lawrence}, {Le Jeune}, {Lemos}, {Lesgourgues}, {Levrier},
  {Lewis}, {Liguori}, {Lilje}, {Lilley}, {Lindholm}, {L{\'o}pez-Caniego},
  {Lubin}, {Ma}, {Mac{\'\i}as-P{\'e}rez}, {Maggio}, {Maino}, {Mandolesi},
  {Mangilli}, {Marcos-Caballero}, {Maris}, {Martin}, {Martinelli},
  {Mart{\'\i}nez-Gonz{\'a}lez}, {Matarrese}, {Mauri}, {McEwen}, {Meinhold},
  {Melchiorri}, {Mennella}, {Migliaccio}, {Millea}, {Mitra},
  {Miville-Desch{\^e}nes}, {Molinari}, {Montier}, {Morgante}, {Moss}, {Natoli},
  {N{\o}rgaard-Nielsen}, {Pagano}, {Paoletti}, {Partridge}, {Patanchon},
  {Peiris}, {Perrotta}, {Pettorino}, {Piacentini}, {Polastri}, {Polenta},
  {Puget}, {Rachen}, {Reinecke}, {Remazeilles}, {Renzi}, {Rocha}, {Rosset},
  {Roudier}, {Rubi{\~n}o-Mart{\'\i}n}, {Ruiz-Granados}, {Salvati}, {Sandri},
  {Savelainen}, {Scott}, {Shellard}, {Sirignano}, {Sirri}, {Spencer},
  {Sunyaev}, {Suur-Uski}, {Tauber}, {Tavagnacco}, {Tenti}, {Toffolatti},
  {Tomasi}, {Trombetti}, {Valenziano}, {Valiviita}, {Van Tent}, {Vibert},
  {Vielva}, {Villa}, {Vittorio}, {Wandelt}, {Wehus}, {White}, {White},
  {Zacchei}, \& {Zonca}}]{planck2020}
{Planck Collaboration}, {Aghanim}, N., {Akrami}, Y., {et~al.} 2020, \aap, 641,
  A6

\bibitem[{{Pranav}(2021)}]{pranav2021}
{Pranav}, P. 2021, arXiv e-prints, arXiv:2109.08721

\bibitem[{{Pranav} {et~al.}(2017){Pranav}, {Edelsbrunner}, {van de Weygaert},
  {Vegter}, {Kerber}, {Jones}, \& {Wintraecken}}]{pranav2017}
{Pranav}, P., {Edelsbrunner}, H., {van de Weygaert}, R., {et~al.} 2017, \mnras,
  465, 4281

\bibitem[{Pun {et~al.}(2018)Pun, Xia, \& Lee}]{PXL:PersistentHomologybased}
Pun, C.~S., Xia, K., \& Lee, S.~X. 2018, arXiv e-prints, arXiv:1811.00252v1

\bibitem[{{Pyne} \& {Joachimi}(2021)}]{Pyne2021}
{Pyne}, S. \& {Joachimi}, B. 2021, \mnras, 503, 2300

\bibitem[{Reininghaus {et~al.}(2015)Reininghaus, Huber, Bauer, \&
  Kwitt}]{RHBK:stablemultiscale}
Reininghaus, J., Huber, S., Bauer, U., \& Kwitt, R. 2015, in Proceedings of the
  IEEE conference on computer vision and pattern recognition, 4741--4748

\bibitem[{{Remus} {et~al.}(2017){Remus}, {Dolag}, \& {Hoffmann}}]{Remus:2017}
{Remus}, R.-S., {Dolag}, K., \& {Hoffmann}, T. 2017, Galaxies, 5, 49

\bibitem[{{Saro} {et~al.}(2014){Saro}, {Liu}, {Mohr}, {Aird}, {Ashby},
  {Bayliss}, {Benson}, {Bleem}, {Bocquet}, {Brodwin}, {Carlstrom}, {Chang},
  {Chiu}, {Cho}, {Clocchiatti}, {Crawford}, {Crites}, {de Haan}, {Desai},
  {Dietrich}, {Dobbs}, {Dolag}, {Dudley}, {Foley}, {Gangkofner}, {George},
  {Gladders}, {Gonzalez}, {Halverson}, {Hennig}, {Hlavacek-Larrondo},
  {Holzapfel}, {Hrubes}, {Jones}, {Keisler}, {Lee}, {Leitch}, {Lueker},
  {Luong-Van}, {Mantz}, {Marrone}, {McDonald}, {McMahon}, {Mehl}, {Meyer},
  {Mocanu}, {Montroy}, {Murray}, {Nurgaliev}, {Padin}, {Patej}, {Pryke},
  {Reichardt}, {Rest}, {Ruel}, {Ruhl}, {Saliwanchik}, {Sayre}, {Schaffer},
  {Shirokoff}, {Spieler}, {Stalder}, {Staniszewski}, {Stark}, {Story}, {van
  Engelen}, {Vanderlinde}, {Vieira}, {Vikhlinin}, {Williamson}, {Zahn}, \&
  {Zenteno}}]{Saro:2014}
{Saro}, A., {Liu}, J., {Mohr}, J.~J., {et~al.} 2014, \mnras, 440, 2610

\bibitem[{{Schirmer} {et~al.}(2007){Schirmer}, {Erben}, {Hetterscheidt}, \&
  {Schneider}}]{Schirmer:2007}
{Schirmer}, M., {Erben}, T., {Hetterscheidt}, M., \& {Schneider}, P. 2007,
  \aap, 462, 875

\bibitem[{{Schneider}(1996)}]{Schneider:1996}
{Schneider}, P. 1996, \mnras, 283, 837

\bibitem[{{Schneider} {et~al.}(2010){Schneider}, {Eifler}, \&
  {Krause}}]{Schneider:2010}
{Schneider}, P., {Eifler}, T., \& {Krause}, E. 2010, \aap, 520, A116

\bibitem[{{Secco} {et~al.}(2022){Secco}, {Samuroff}, {Krause}, {Jain},
  {Blazek}, {Raveri}, {Campos}, {Amon}, {Chen}, {Doux}, {Choi}, {Gruen},
  {Bernstein}, {Chang}, {DeRose}, {Myles}, {Fert{\'e}}, {Lemos}, {Huterer},
  {Prat}, {Troxel}, {MacCrann}, {Liddle}, {Kacprzak}, {Fang}, {S{\'a}nchez},
  {Pandey}, {Dodelson}, {Chintalapati}, {Hoffmann}, {Alarcon}, {Alves},
  {Andrade-Oliveira}, {Baxter}, {Bechtol}, {Becker}, {Brandao-Souza},
  {Camacho}, {Carnero Rosell}, {Carrasco Kind}, {Cawthon}, {Cordero}, {Crocce},
  {Davis}, {Di Valentino}, {Drlica-Wagner}, {Eckert}, {Eifler}, {Elidaiana},
  {Elsner}, {Elvin-Poole}, {Everett}, {Fosalba}, {Friedrich}, {Gatti},
  {Giannini}, {Gruendl}, {Harrison}, {Hartley}, {Herner}, {Huang}, {Huff},
  {Jarvis}, {Jeffrey}, {Kuropatkin}, {Leget}, {Muir}, {Mccullough}, {Navarro
  Alsina}, {Omori}, {Park}, {Porredon}, {Rollins}, {Roodman}, {Rosenfeld},
  {Ross}, {Rykoff}, {Sanchez}, {Sevilla-Noarbe}, {Sheldon}, {Shin}, {Tutusaus},
  {Varga}, {Weaverdyck}, {Wechsler}, {Yanny}, {Yin}, {Zhang}, {Zuntz},
  {Abbott}, {Aguena}, {Allam}, {Annis}, {Bacon}, {Bertin}, {Bhargava},
  {Bridle}, {Brooks}, {Buckley-Geer}, {Burke}, {Carretero}, {Costanzi}, {da
  Costa}, {De Vicente}, {Diehl}, {Dietrich}, {Doel}, {Ferrero}, {Flaugher},
  {Frieman}, {Garc{\'\i}a-Bellido}, {Gaztanaga}, {Gerdes}, {Giannantonio},
  {Gschwend}, {Gutierrez}, {Hinton}, {Hollowood}, {Honscheid}, {Hoyle},
  {James}, {Jeltema}, {Kuehn}, {Lahav}, {Lima}, {Lin}, {Maia}, {Marshall},
  {Martini}, {Melchior}, {Menanteau}, {Miquel}, {Mohr}, {Morgan}, {Ogando},
  {Palmese}, {Paz-Chinch{\'o}n}, {Petravick}, {Pieres}, {Plazas Malag{\'o}n},
  {Rodriguez-Monroy}, {Romer}, {Sanchez}, {Scarpine}, {Schubnell}, {Scolnic},
  {Serrano}, {Smith}, {Soares-Santos}, {Suchyta}, {Swanson}, {Tarle}, {Thomas},
  {To}, \& {DES Collaboration}}]{DESY3_Secco}
{Secco}, L.~F., {Samuroff}, S., {Krause}, E., {et~al.} 2022, \prd, 105, 023515

\bibitem[{{Sellentin} \& {Heavens}(2016)}]{Sellentin:2016}
{Sellentin}, E. \& {Heavens}, A.~F. 2016, \mnras, 456, L132

\bibitem[{{Semboloni} {et~al.}(2011){Semboloni}, {Hoekstra}, {Schaye}, {van
  Daalen}, \& {McCarthy}}]{Semboloni11}
{Semboloni}, E., {Hoekstra}, H., {Schaye}, J., {van Daalen}, M.~P., \&
  {McCarthy}, I.~G. 2011, \mnras, 417, 2020

\bibitem[{{Sheldon} \& {Huff}(2017)}]{METACAL}
{Sheldon}, E.~S. \& {Huff}, E.~M. 2017, \apj, 841, 24

\bibitem[{{Shirasaki} \& {Yoshida}(2014)}]{shirasaki2014}
{Shirasaki}, M. \& {Yoshida}, N. 2014, \apj, 786, 43

\bibitem[{{Sousbie}(2011)}]{sousbie2011}
{Sousbie}, T. 2011, \mnras, 414, 350

\bibitem[{{Spergel} {et~al.}(2013){Spergel}, {Gehrels}, {Breckinridge},
  {Donahue}, {Dressler}, {Gaudi}, {Greene}, {Guyon}, {Hirata}, {Kalirai},
  {Kasdin}, {Moos}, {Perlmutter}, {Postman}, {Rauscher}, {Rhodes}, {Wang},
  {Weinberg}, {Centrella}, {Traub}, {Baltay}, {Colbert}, {Bennett},
  {Kiessling}, {Macintosh}, {Merten}, {Mortonson}, {Penny}, {Rozo},
  {Savransky}, {Stapelfeldt}, {Zu}, {Baker}, {Cheng}, {Content}, {Dooley},
  {Foote}, {Goullioud}, {Grady}, {Jackson}, {Kruk}, {Levine}, {Melton},
  {Peddie}, {Ruffa}, \& {Shaklan}}]{spergel2013}
{Spergel}, D., {Gehrels}, N., {Breckinridge}, J., {et~al.} 2013, arXiv
  e-prints, arXiv:1305.5422

\bibitem[{{Steinborn} {et~al.}(2016){Steinborn}, {Dolag}, {Comerford},
  {Hirschmann}, {Remus}, \& {Teklu}}]{Steinborn:2016}
{Steinborn}, L.~K., {Dolag}, K., {Comerford}, J.~M., {et~al.} 2016, \mnras,
  458, 1013

\bibitem[{{Steinborn} {et~al.}(2015){Steinborn}, {Dolag}, {Hirschmann},
  {Prieto}, \& {Remus}}]{Steinborn:2015}
{Steinborn}, L.~K., {Dolag}, K., {Hirschmann}, M., {Prieto}, M.~A., \& {Remus},
  R.-S. 2015, \mnras, 448, 1504

\bibitem[{{Takahashi} {et~al.}(2012){Takahashi}, {Sato}, {Nishimichi},
  {Taruya}, \& {Oguri}}]{Takahashi2012}
{Takahashi}, R., {Sato}, M., {Nishimichi}, T., {Taruya}, A., \& {Oguri}, M.
  2012, \apj, 761, 152

\bibitem[{{Teklu} {et~al.}(2015){Teklu}, {Remus}, {Dolag}, {Beck}, {Burkert},
  {Schmidt}, {Schulze}, \& {Steinborn}}]{Teklu:2015}
{Teklu}, A.~F., {Remus}, R.-S., {Dolag}, K., {et~al.} 2015, \apj, 812, 29

\bibitem[{{Troxel} {et~al.}(2018){Troxel}, {MacCrann}, {Zuntz}, {Eifler},
  {Krause}, {Dodelson}, {Gruen}, {Blazek}, {Friedrich}, {Samuroff}, {Prat},
  {Secco}, {Davis}, {Fert{\'e}}, {DeRose}, {Alarcon}, {Amara}, {Baxter},
  {Becker}, {Bernstein}, {Bridle}, {Cawthon}, {Chang}, {Choi}, {De Vicente},
  {Drlica-Wagner}, {Elvin-Poole}, {Frieman}, {Gatti}, {Hartley}, {Honscheid},
  {Hoyle}, {Huff}, {Huterer}, {Jain}, {Jarvis}, {Kacprzak}, {Kirk}, {Kokron},
  {Krawiec}, {Lahav}, {Liddle}, {Peacock}, {Rau}, {Refregier}, {Rollins},
  {Rozo}, {Rykoff}, {S{\'a}nchez}, {Sevilla-Noarbe}, {Sheldon}, {Stebbins},
  {Varga}, {Vielzeuf}, {Wang}, {Wechsler}, {Yanny}, {Abbott}, {Abdalla},
  {Allam}, {Annis}, {Bechtol}, {Benoit-L{\'e}vy}, {Bertin}, {Brooks},
  {Buckley-Geer}, {Burke}, {Carnero Rosell}, {Carrasco Kind}, {Carretero},
  {Castander}, {Crocce}, {Cunha}, {D'Andrea}, {da Costa}, {DePoy}, {Desai},
  {Diehl}, {Dietrich}, {Doel}, {Fernandez}, {Flaugher}, {Fosalba},
  {Garc{\'\i}a-Bellido}, {Gaztanaga}, {Gerdes}, {Giannantonio}, {Goldstein},
  {Gruendl}, {Gschwend}, {Gutierrez}, {James}, {Jeltema}, {Johnson}, {Johnson},
  {Kent}, {Kuehn}, {Kuhlmann}, {Kuropatkin}, {Li}, {Lima}, {Lin}, {Maia},
  {March}, {Marshall}, {Martini}, {Melchior}, {Menanteau}, {Miquel}, {Mohr},
  {Neilsen}, {Nichol}, {Nord}, {Petravick}, {Plazas}, {Romer}, {Roodman},
  {Sako}, {Sanchez}, {Scarpine}, {Schindler}, {Schubnell}, {Smith}, {Smith},
  {Soares-Santos}, {Sobreira}, {Suchyta}, {Swanson}, {Tarle}, {Thomas},
  {Tucker}, {Vikram}, {Walker}, {Weller}, {Zhang}, \& {DES
  Collaboration}}]{troxel2018}
{Troxel}, M.~A., {MacCrann}, N., {Zuntz}, J., {et~al.} 2018, \prd, 98, 043528

\bibitem[{{Uzeirbegovic} {et~al.}(2020){Uzeirbegovic}, {Geach}, \&
  {Kaviraj}}]{arXiv:2004.06734}
{Uzeirbegovic}, E., {Geach}, J.~E., \& {Kaviraj}, S. 2020, \mnras, 498, 4021

\bibitem[{{van Daalen} {et~al.}(2014){van Daalen}, {Schaye}, {McCarthy},
  {Booth}, \& {Dalla Vecchia}}]{OWLS}
{van Daalen}, M.~P., {Schaye}, J., {McCarthy}, I.~G., {Booth}, C.~M., \& {Dalla
  Vecchia}, C. 2014, \mnras, 440, 2997

\bibitem[{{van de Weygaert} {et~al.}(2013){van de Weygaert}, {Vegter},
  {Edelsbrunner}, {Jones}, {Pranav}, {Park}, {Hellwing}, {Eldering},
  {Kruithof}, {Bos}, {Hidding}, {Feldbrugge}, {ten Have}, {van Engelen},
  {Caroli}, \& {Teillaud}}]{weygeart2011}
{van de Weygaert}, R., {Vegter}, G., {Edelsbrunner}, H., {et~al.} 2013, arXiv
  e-prints, arXiv:1306.3640

\bibitem[{{Wasserman}(2018)}]{wassermann2018}
{Wasserman}, L. 2018, Annual Review of Statistics and Its Application, 5, 501

\bibitem[{{Xu} {et~al.}(2019){Xu}, {Cisewski-Kehe}, {Green}, \&
  {Nagai}}]{xu2019}
{Xu}, X., {Cisewski-Kehe}, J., {Green}, S.~B., \& {Nagai}, D. 2019, Astronomy
  and Computing, 27, 34

\bibitem[{{Zuntz} {et~al.}(2015){Zuntz}, {Paterno}, {Jennings}, {Rudd},
  {Manzotti}, {Dodelson}, {Bridle}, {Sehrish}, \&
  {Kowalkowski}}]{arXiv:1409.3409}
{Zuntz}, J., {Paterno}, M., {Jennings}, E., {et~al.} 2015, Astronomy and
  Computing, 12, 45

\bibitem[{{Z{\"u}rcher} {et~al.}(2021){Z{\"u}rcher}, {Fluri}, {Sgier},
  {Kacprzak}, \& {Refregier}}]{zurcher2021}
{Z{\"u}rcher}, D., {Fluri}, J., {Sgier}, R., {Kacprzak}, T., \& {Refregier}, A.
  2021, \jcap, 2021, 028

\end{thebibliography}

\begin{appendix}
\section{Data compression of Heatmaps}
\label{sec:app_data_compression}
\begin{figure}
    % \sidecaption
    \resizebox{\hsize}{!}{\includegraphics{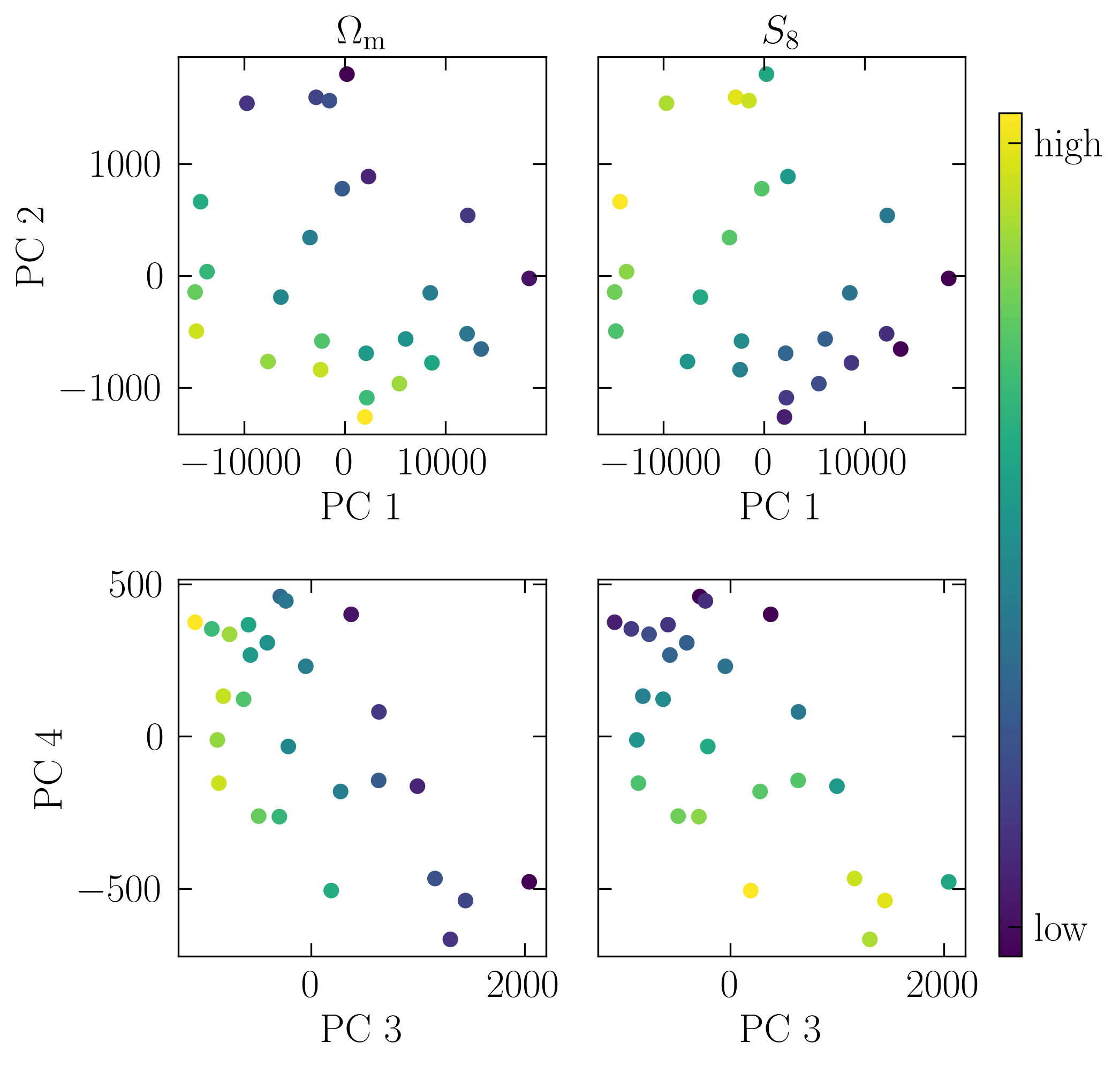}}
    \caption{The cosmology dependence of the first four principal components. Each point in the scatter plot represents one of the 26 cosmologies of the {\it cosmo}-SLICS. In the top row, the $x$- and $y$-coordinates correspond to the value of the first two principle components, in the bottom row they correspond to the values of the third and fourth principal component. In the left column, the colours represent the value of $\Omm$ of the respective {\it cosmo}-SLICS simulation, in the right column they denote the value of $S_8$.}
    \label{fig:components_PCA}
\end{figure}

\begin{figure*}
    \includegraphics[width=17cm]{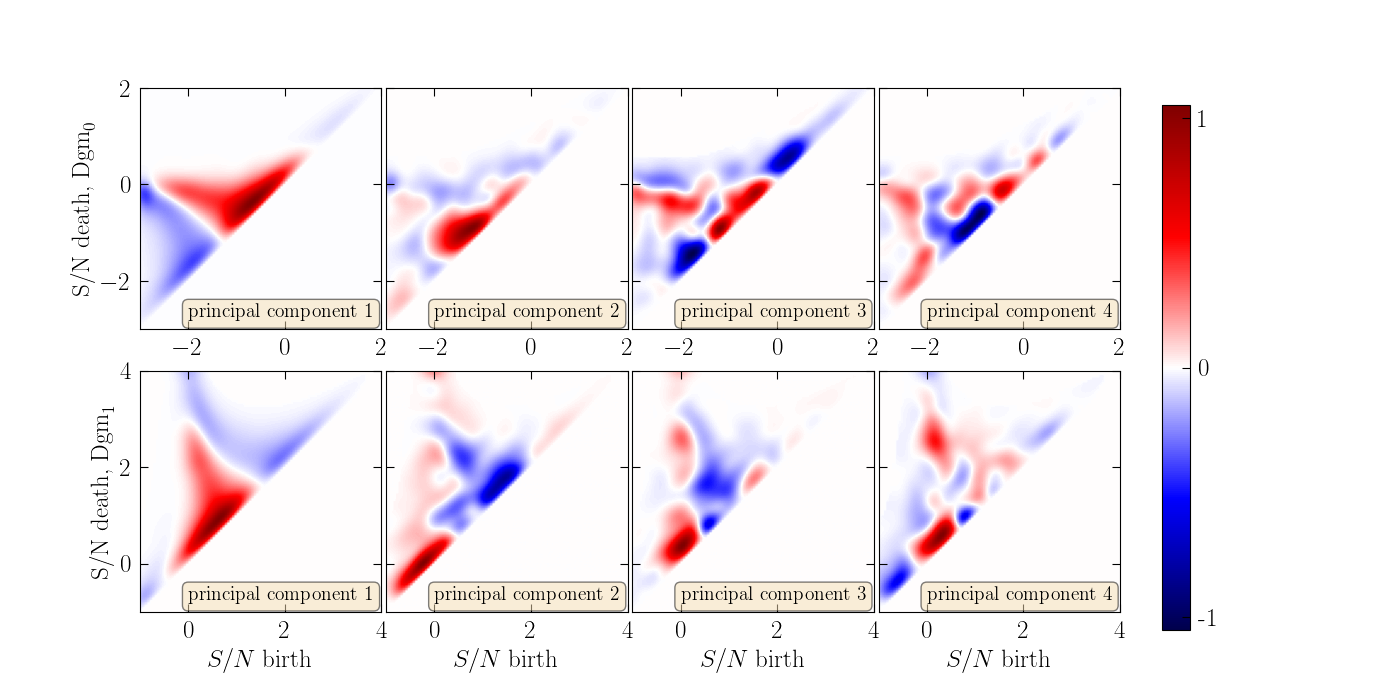}
    \caption{The first four (normalised) principal components of the heatmaps in a principal component analysis.}
    \label{fig:principal_components}
\end{figure*}
As the raw heatmaps contain 10100 entries per combination of tomographic redshift bins, a direct cosmological parameter analysis with these maps is currently impossible. We therefore need to explore different methods of compressing the raw data.

Our first approach at data compression is a principal component analysis (PCA). This rather simple method is highly efficient at reducing complex data to only a few manageable dimensions \citep[see e.g.][]{arXiv:2004.06734}. For each combination of tomographic redshift bins, we apply a PCA to the heatmap extracted from all 19 regions. We see that the PCA correctly identifies that the differences between the S/N maps of the different {\it cosmo}-SLICS are driven by the changes of the cosmological parameters $\Omm$ and $S_8$.

A PCA is still an incredibly useful tool to not only extract cosmological information from a data vector, but also to understand the behaviour of the data itself better. For example, Fig.~\ref{fig:components_PCA} shows that the first principal component is almost exactly antiproportional to $S_8$, whereas the second principal component is proportional to $\Omm$. Comparing these findings with Fig.~\ref{fig:principal_components}, we see that a high value of $\Omm$ leads to a large number of features in $\dgm_1$ (peaks) being born and dying between signal-to-noise values of 1 and 2, whereas a low value of $\Omm$ leads to more features in $\dgm_1$ being born and dying between S/N values of $-0.5$ and $0.5$. A similar analysis for the first principal components yields the expected conclusion that a higher value of $S_8$ leads to more peaks being born and dying at higher S/N values, and more voids being born and dying at lower S/N values.

One disadvantage of PCA is the fact that it is not straightforward to include the internal covariance of the data vector: While the PCA might detect huge differences between two different {\it cosmo}-SLICS simulations, these might just be caused by the fact that this specific part of the data vector is particularly noisy, and not by differences in the cosmological signal.
%Other data compression methods like MOPED or \citet{Asgari:2015} do take the sample covariance into account, but they do require knowledge about the inverse covariance matrix, which can only be computed if the number of simulations used to determine the covariance is greater than the dimensions of the data vector \cite{Hartlap:2007}. 

A more sophisticated method of data compression is the Massively
Optimized Parameter Estimation and Data compression \citep[MOPED,][]{Heavens:2000,Heavens:2017,Ferreira:2021}. Assuming a Gaussian likelihood, Gaussian posterior distributions and a constant covariance matrix $C$, this compression method preserves the entirety of the Fisher information to $N_\mathrm{param}$ dimensions, where $N_\mathrm{param}$ is the number of cosmological (and nuisance) parameters present in the inference. However, this method uses the Fisher formalism, and thus knowledge of the inverse covariance matrix $C^{-1}$ is required. As our uncompressed data vector contains $151\,500$ entries and we can only estimate $C$ with about $10^3$ sets of simulations, the matrix is singular and thus not invertible \citep{Hartlap:2007}. We therefore opted for sub-sampling our data vector and performing a MOPED compression for each individual combination of tomographic redshift bins, but doing this we neglected the information contained in the cross-correlation between different combinations of redshift bins, yielding parameter constraints that were not competitive with the ones from other data compression methods.

While MOPED is an elegant method to compress a data vector to the absolute minimum of required dimensions, this also means that all additional information that was not part of this data compression gets lost. In particular, imperfect knowledge of the covariance matrix and noisy derivatives heavily affect the constraining power of MOPED. \citet{Asgari:2015} analysed this loss of information and developed a method that is more stable with respect to changes in the covariance matrix and derivatives and offers more constraining power in the case of non-linear parameter degeneracies (like the one between $\Omm$ and $\sigma_8$).

All things considered, all data compression methods manage to extract a comparable amount of information out of the raw data vector (see Fig.~\ref{fig:data_compression_comparison}). We thus opt for the $\chi^2$-maximiser method as the data vector obtained from this method is easiest to interpret.

\section{On the observed $\Omm$ tension in the analysis of DES-Y1}
\label{sec:app_tension}
When analysing DES-Y1 data, we observe a 3.2$\sigma$ tension between the values of $\Omm$ estimated from 2PCF and persistent homology. In principle, there are several possible scenarios that can cause this tension. Our validation tests show that this is unlikely to be caused by a bug in the pipeline. A second possibility is that this is caused by a statistical fluctuation in the data. The third, and most interesting, scenario would be the presence of something unknown (and thus unaccounted for) in the DES-Y1 data. This could either be a systematic effect that we have not properly taken into account, or a sign of deviations from the $w$CDM cosmological model that affects the topological structure of the data, but not its two-point statistics. For example, we know that persistent homology is very sensitive to primordial non-Gaussianities in the large-scale structure \citep{Biagetti:2022}, which can not be detected by two-point statistics.

\begin{figure}
    \resizebox{\hsize}{!}{\includegraphics{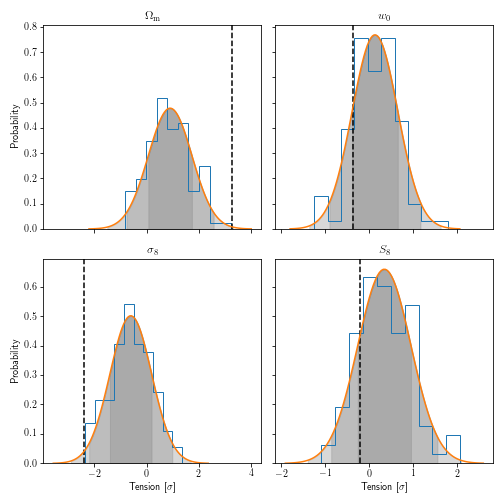}}
    \caption{A histogram of the tensions in cosmological parameters between 2PCF and persistent homology measured on 100 individual lines-of-sights in the {\it Covariance Training Set} (blue) and a Gaussian fit to these values (orange). The actual tension we measured in DES-Y1 is shown by the dashed black line. No tension is observed when running our pipeline on the mock data vector constructed from all simulations of the {\it Covariance Training Set}, which is about $12\times$ larger.}
    \label{fig:cosmo_params_tension}
\end{figure}

\begin{figure*}
    \centering
    \sidecaption
    \includegraphics[width=12cm]{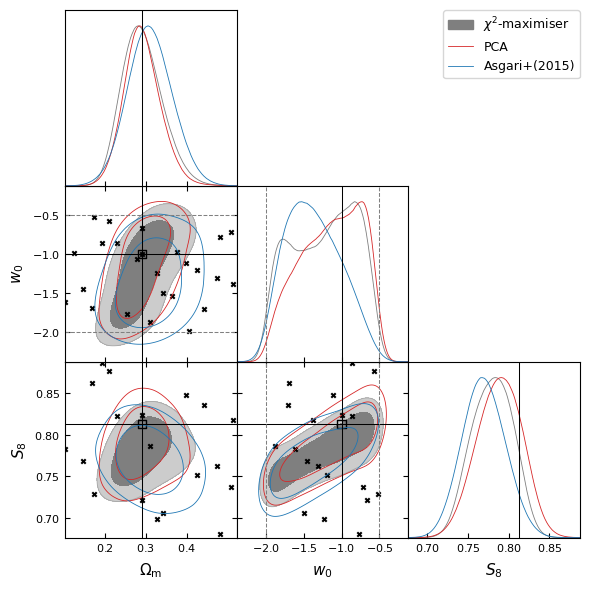}
    \caption{A comparison of the constraining power of different data compression methods. Our chosen method, the $\chi^2$-maximiser is shown in grey, two alternative methods \citep[PCA and][in red and blue, respectively]{Asgari:2015}.}
    \label{fig:data_compression_comparison}
\end{figure*}

The fact that we achieve extraordinarily consistent results with HD+21 using a fully independent measurement and inference pipeline points to the conclusion that this tension is not caused by a bug in the pipeline. To investigate the probability of a statistical fluctuation causing this effect, we run our inference pipeline for both 2PCF and persistent homology on 100 individual lines-of-sight of the {\it Covariance Training Sample}. For each individual line-of-sight, we then estimate the tension between 2PCF and persistent homology on each cosmological parameter. The results are shown in Fig.~\ref{fig:cosmo_params_tension}. We observe that persistent homology seems to favor higher values of $\Omm$ and lower values of $\sigma_8$ than 2PCF, while $S_8$ remains relatively unbiased. We assume that these tensions follow a normal distribution and compute its mean and variance, constructing a Gaussian fit to the values. According to this analysis, the chance that the observed bias in $\Omm$ is due to a statistical fluctuation is at $0.5\%$ ($2.6\%$ for $\sigma_8$), which is still unlikely, but not as unlikely as the initial $3.2\sigma$ tension we observed suggests. Recall that no tension is observed when running our pipeline on the mock data vector constructed from all simulations of the {\it Covariance Training Set}, which is about $12\times$ larger, suggesting that the observed tension results from statistical fluctuations that are averaged down in our validation test.

This effect certainly warrants further investigation. If something similar shows up in an analysis of KiDS-1000 (Harnois-D\'eraps et al.~in prep., Heydenreich et al.~in prep.), an investigation into potential causes for a bias in $\Omm$ becomes highly warranted. If, however, that analysis does not show any bias in $\Omm$, we can assume that this tension is likely to be a mere statistical fluctuation in the data. 

\section{Complete parameter constraints of the MCMC}

% \begin{figure*}
%     \sidecaption
%     \includegraphics[width=12cm]{components_MOPED.png}
%     \caption{The cosmology dependency of the first MOPED. Each point in the scatter plot represents one of the 26 cosmologies of the cosmoSLICS. The $x$- and $y$-coordinates correspond to the value of the first two principle components. In the left column, the colours represent the value of $\Omm$ of the respective cosmoSLICS simulation, in the right column they denote the value of $S_8$.}
%     \label{fig:components_PCA}
% \end{figure*}

% MOPED manages a cleaner separation between the different components and cosmological parameters
% \section{Nuisance parameters and priors}

% \JHD{[]Add a few words about this appendix here]}

% \begin{figure*}
%     \centering
%     \includegraphics[width=17cm]{marginalisations_all_params.png}
%     \caption{The full likelihood results of the analysis described in Fig.~\ref{fig:marginalisation_systematics}}
%     \label{fig:marginalisation_systematics_full}
% \end{figure*}

\begin{figure*}
    \centering
    \includegraphics[width=17cm]{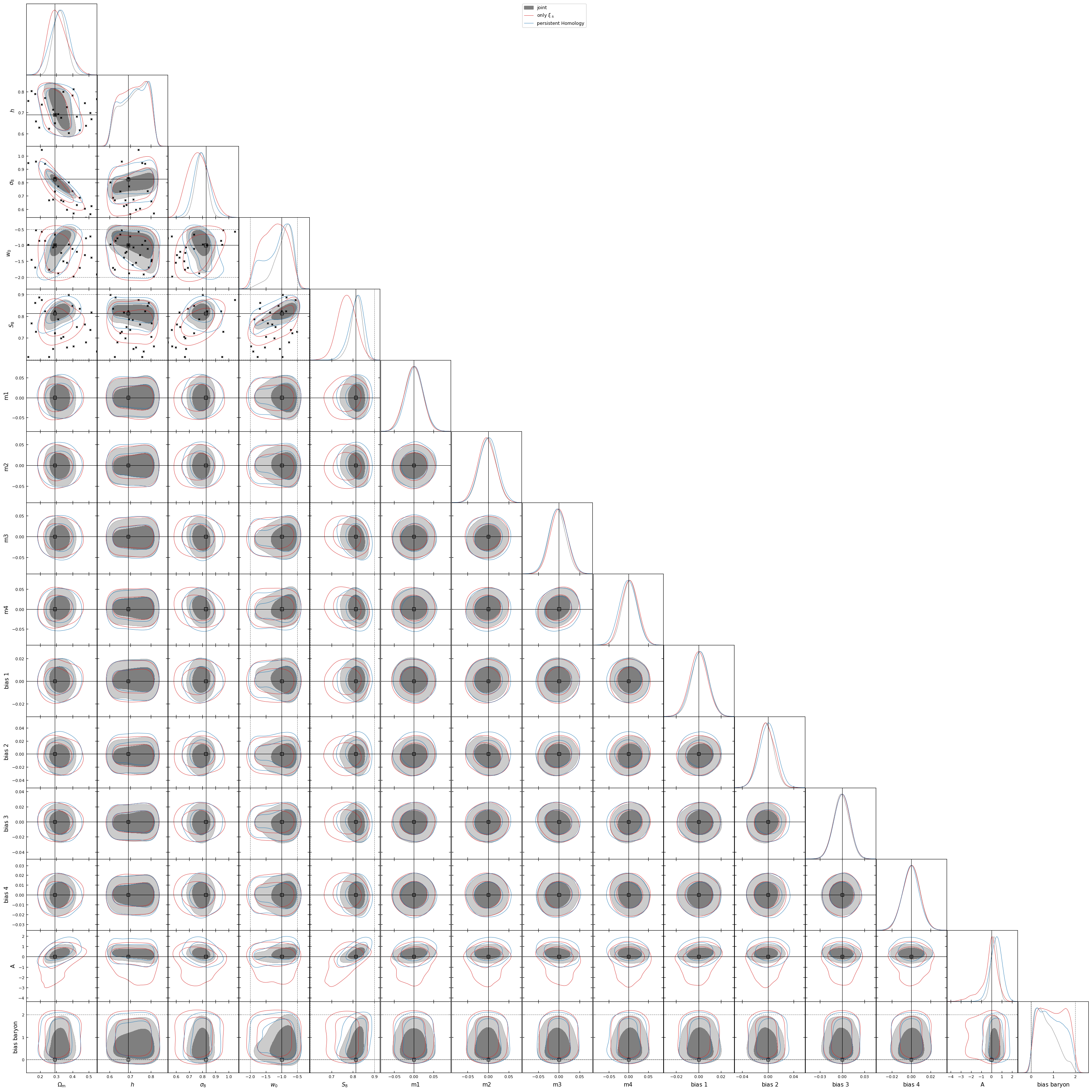}
    \caption{Same as Fig.\ref{fig:mcmc_joint} with all cosmological and nuisance parameters.}
    \label{fig:mcmc_joint_full}
\end{figure*}

\begin{figure*}
    \centering
    \includegraphics[width=17cm]{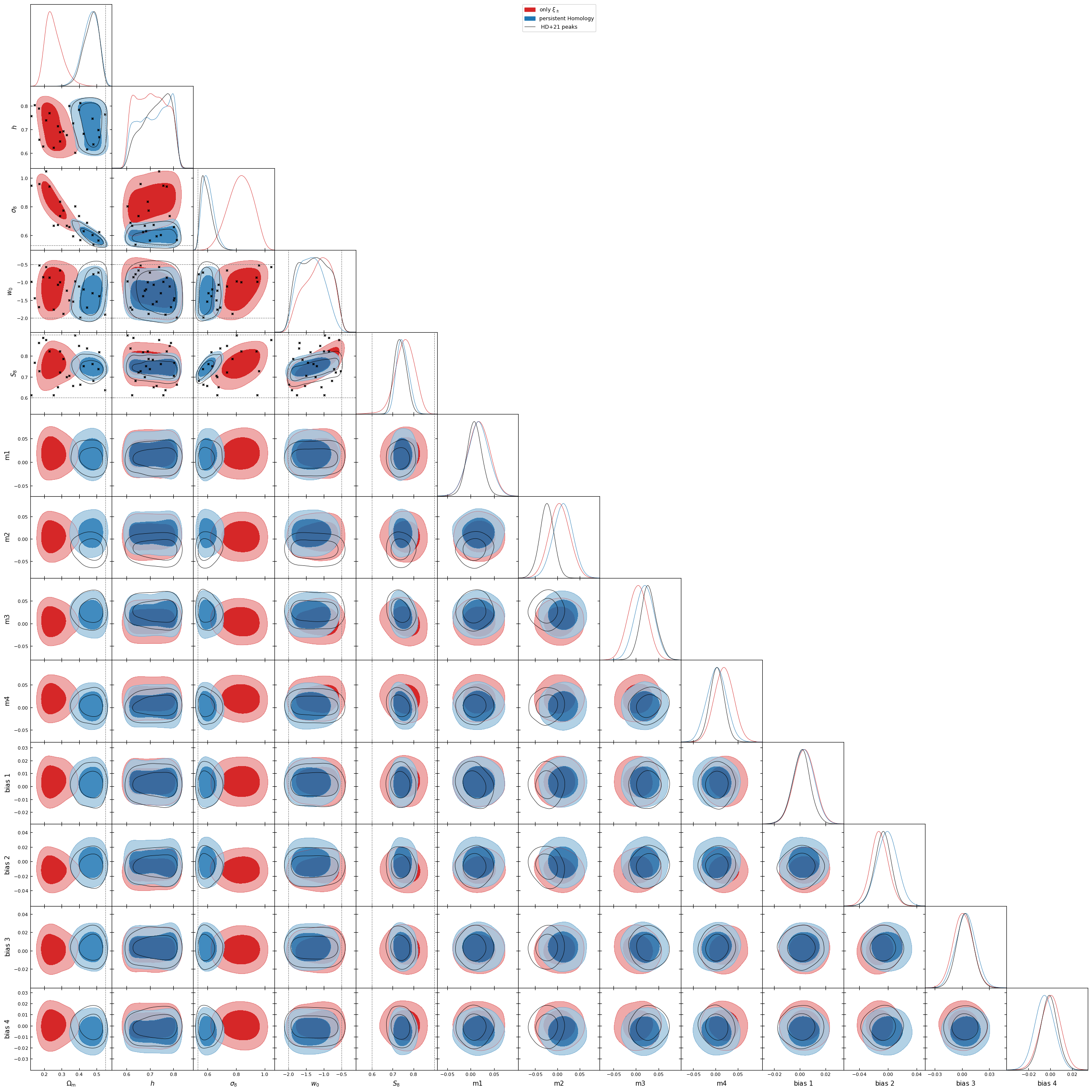}
    \caption{Same as Fig.\ref{fig:mcmc_joint_des} with all cosmological and nuisance parameters.}
    \label{fig:mcmc_joint_des_full}
\end{figure*}

\end{appendix}

\end{document}